\documentclass[preprint,12pt,3p]{elsarticle}

\usepackage{graphicx}
\usepackage{caption}
\usepackage{subcaption}
\usepackage{booktabs}
\usepackage{amssymb}
\usepackage{amsthm,amsmath}
\usepackage[numbers]{natbib}

\journal{J. Comp. Phys.}

\begin{document}

\begin{frontmatter}

\title{Implementation of a hybrid particle code with a PIC description  in r-z and a gridless description in $\phi$ into OSIRIS}

\author[label1]{A. Davidson}
\address[label1]{Department of Physics and Astronomy, University of California, Los Angeles, California 90095, USA}


\ead{davidsoa@physics.ucla.edu}

\address[label2]{Department of Electrical Engineering, University of California, Los Angeles, California 90095, USA}

\author[label1]{A. Tableman}
\ead{Tableman@physics.ucla.edu}

\author[label1]{W. An}
\ead{anweiming@ucla.edu}

\author[label1]{F.S. Tsung}
\ead{tsung@physics.ucla.edu}

\author[label1]{W. Lu}
\ead{luwei@ucla.edu}

\author[label3]{J. Vieira}
\ead{jorge.vieira@ist.utl.pt}

\author[label3,label4]{R.A. Fonseca}
\ead{ricardo.fonseca@iscte.pt}

\author[label3]{L.O. Silva}
\ead{luis.silva@ist.utl.pt}

\author[label1,label2]{W.B. Mori}
\ead{mori@physics.ucla.edu}
\address[label3]{GoLP/Instituto de Plasmas e Fusão Nuclear, Instituto Superior T\'ecnico, Universidade de Lisboa, Lisbon, Portugal}
\address[label4]{Departamento Ciências e Tecnologias da Informação, ISCTE-Instituto Universitário de Lisboa, 1649-026 Lisboa, Portugal}

\begin{abstract}
For many plasma physics problems, three-dimensional and kinetic effects are very important. However, such simulations are very computationally intensive. Fortunately, there is a class of problems for which there is nearly azimuthal symmetry and the dominant three-dimensional physics is captured by the inclusion of only a few azimuthal harmonics. Recently, it was proposed \cite{Lifschitz} to model one such problem, laser wakefield acceleration, by expanding the fields and currents in azimuthal harmonics and truncating the expansion after only the first harmonic. The complex amplitudes of the fundamental and first harmonic for the fields were solved on an r-z grid and a procedure for calculating the complex current amplitudes for each particle based on its motion in Cartesian geometry was presented using a Marder's correction to maintain the validity of Gauss's law. In this paper, we describe an implementation of this algorithm into OSIRIS  using a rigorous charge conserving current deposition method to maintain the validity of Gauss's law. We show that this algorithm is a hybrid method which uses a particles-in-cell description in r-z and a gridless description in $\phi$. We include the ability to keep an arbitrary number of harmonics and higher order particle shapes. Examples, for laser wakefield acceleration, plasma wakefield acceleration, and beam loading are also presented and directions for future work are discussed. 

\end{abstract}


\end{frontmatter}


\section{Introduction}
\label{introduction}

Particle-in-cell simulations are widely used and well established for simulating plasmas in fields  ranging from magnetic fusion, inertial confinement fusion, plasma based acceleration, and space and astrophysics. These simulations are conducted in one, two, and three dimensions. The two dimensional simulations are often conducted in cartesian "slab" geometry or r-z "cylindrical" geometry. While the two dimensional simulations can be very useful for carrying out parameter scans and illuminating physics , there are some problems in which three dimensional effects lead to both qualitative or quantitative differences. For example,  in plasma based acceleration the space charge forces of an  intense particle beam or the radiation pressure of an intense laser, drive plasma wave wake fields as they propagate through long regions of plasmas. The wakefield structure, the self-trapping of electrons, the beam loading of wakes by trailing beams, and the evolution of the drive particle or laser beams are not properly modeled in 2D slab geometry due to geometrical effects. Therefore r-z PIC simulations have been used to properly describe the structure of the wakefield. However, the use of an r-z code precludes hosing and the effect of asymmetric spot sizes of both the drive and trailing particle beams. In addition, a linearly polarized (or circularly polarized) laser is not typically azimuthally symmetric (azimuthally symmetric laser pulses are radially polarized) so a laser driver cannot even be modeled using an r-z code. 

Several methods have been developed for more efficiently modeling plasma-based acceleration in three dimensions (or in lower dimensions). These include the moving window method \cite{Decker94}, quasi-static methods \cite{MoraAntonsen97, Huang06,an:13}, the ponderomotive guiding center (PGC) method for modeling laser propagation \cite{MoraAntonsen97,Gordon00}, and the use of simulating the physics in Lorentz boosted frames \cite{Vay07,Vay11,Martins10,Martins09}. In some cases these methods are combined. For example, a combination of quasi-static field equations and the ponderomotive guiding center approximation are used in QuickPIC\cite{an:13,Huang06} to model laser wakefield acceleration. 
In addition, each of these methods have advantages and disadvantages when compared to full PIC methods in the rest frame of the plasma. The quasi-static methods cannot accurately model self-injection, the ponderomotive guiding center cannot model full pump depletion distances for extremely high laser intensities, and the Lorentz boosted frame method still has issues with numerical Cherenkov instabilities and it has not been well tested for studying self-trapping where particle statistics can be an issue \cite{xu13, yu14, Godfrey13} . Work continues on each of these methods.

Very recently, an algorithm was proposed that would allow modeling laser propagation with similar computational costs to an r-z code. In this algorithm the fields and currents are expanded into azimuthal harmonics (modes) where the amplitudes of each harmonic are complex and functions of r and z. This expansion is substituted into Maxwell's equations to generate a series of equations for the complex amplitudes for each harmonic. In \cite{Lifschitz} the expansion was truncated after the first harmonic. The particles are pushed in 3D cartesian geometry and are then used to obtain the complex amplitudes for each harmonic of the current. In \cite{Lifschitz}~the current deposition method did not conserve charge so a Marder's method \cite{Marder} was used to maintain the accuracy of Gauss's law. The Marder's method is an approximation to the Boris correction \cite{Langdon92,birdsall1985plasma} in which  a correction, $\mathbf{E_c}$ is added to an uncorrected field, $\mathbf{E^{'}}$ such that $\nabla \cdot (\mathbf{E^{'}} + \mathbf{E}_c) = \rho$. The correction to the field is defined as $\mathbf{E_c} =-\nabla \phi_c$ where $\nabla^2\phi_c=\nabla \cdot(\mathbf{E})-\rho$. They also showed results for laser wakefield acceleration and found agreement with a full PIC code.

In this paper, we describe the implementation of such a truncated azimuthal Fourier decomposition (i.e., harmonic expansion) into the OSIRIS simulation framework. OSIRIS is a fully parallelized PIC finite-difference code that has been used in 1D, 2D, 3D geometries\cite{ricardo}. For 2D simulations a cartesian slab (xz) or a cylindrical (rz) geometry can be used. We reused as much of the existing 2D r-z structure as possible. We view this algorithm as a hybrid between a traditional PIC method where quantities are defined on an r-z grid and a gridless method\cite{dawson70} in $\phi$ where quantities are expanded in global basis functions (e.g., Fourier modes) defined at all locations and the expansion is truncated. This strategy of combining gridded and gridless algorithms is actually not new. 
For example, in the the early $1980$s Godfrey and collaborators developed  IPROP\cite{iprop, ivory}, which was capable of following an arbitrary number of azimuthal modes to study filamentation as well as hosing of high current electron beams propagating in the atmosphere.

In the implementation for OSIRIS an arbitrary number of harmonics can be kept. 
In addition, OSIRIS uses a rigorous charge conserving current deposition for the PIC part.  Therefore, we have used this as a starting point to develop a current deposition scheme which conserves charge for each harmonic particle by particle. OSIRIS can also use higher order particle shapes so we have implemented this into the PIC part of the algorithm. For the gridless part we have used point particle shapes but have described how to extend this to higher order particle shapes. In addition, OSIRIS can model plasma wakefield acceleration and beam loading. We therefore give examples of such simulations using the new algorithm. We also note that this algorithm could be combined with the PGC as well as Lorentz boosted frame ideas for even more dramatic speed ups over full 3D simulations.

In Section~\ref{theory} we will discuss the mathematical description of Maxwell's equations using an azimuthal harmonic expansion for the electromagnetic fields and currents. In Section~\ref{algorithm} we discuss the specific numerical implementation of these equations, as well as 
the complications which need to be considered in the cell closest to the cylindrical axis.
We also derive the charge conserving current deposition algorithm and discuss its implementation.  In Section ~\ref{results} we  
give examples from the code of laser wakefield acceleration, plasma wakefield acceleration and beam loading in laser driven wakes including comparison with full 3D simulations. We also test the charge conservation and accuracy of Gauss's law for one test case. Last,
in Section~\ref{conclusion}, conclusions and directions for future work are presented.


\section{Theory}
\label{theory}

\subsection{Electromagnetic fields expressed in azimuthal harmonics}
We begin by expanding the electromagnetic fields and the charge ($\rho$) and current densities ($\mathbf{J}$), expressed in cylindrical coordinates, into a Fourier series in $\phi$,
\begin{align}
\mathbf{F}(r,z,\phi) &= \Re\left\{ \sum_{m=0}\mathbf{F}^m(r,z) e^{ i m \phi}\right\} \label{expansion} \\
&= \mathbf{F}^0(r,z) + \Re\{ \mathbf{F}^1\} \cos(\phi) - \Im\{ \mathbf{F}^1\} \sin(\phi) \\
  &~~~~~~~~~~~~~~+ \Re\{ \mathbf{F}^2\} \cos(2\phi) - \Im\{ \mathbf{F}^2\} \sin(2\phi) \nonumber\\
  &~~~~~~~~~~~~~~+ \cdots \nonumber.
\end{align}

Note that the amplitudes of each Fourier harmonic (for all fields) $\mathbf{F}^m$ are complex, whereas the physical fields they are describing, $\mathbf{F}$, are real. As shown in \cite{Lifschitz} a major advantage of this expansion is that it allows modeling a linearly polarized laser with only the first harmonic. Consider a laser with a  polarization angle $\phi_0$ and amplitude $E_0$,
\begin{align}
\label{linlaser}
\mathbf{E}(r,z,\phi) &= E_0 \cos(k_z z - \omega t) \cos(\phi_0) \hat{x} + E_0 \cos(k_z z - \omega_0) \sin(\phi_0) \hat{y}\\
\mathbf{B}(r,z,\phi) &= -E_0 \cos(k_z z - \omega t) \sin(\phi_0) \hat{x} + E_0 \cos(k_z z - \omega_0) \cos(\phi_0) \hat{y},
\end{align}
and let $a(r,z) = E_0 \cos(k_z z - \omega t)$. Decomposing the cartesian unit vectors into cylindrical coordinates unit vectors then gives the radial and azimuthal field components which will have $\sin(\phi)$ and $\cos(\phi)$ terms. By equating these fields to the expansion in   Equation~\ref{expansion}, it can be easily shown that these fields are represented by the $m = 1$ terms 
\begin{align}
E_r^1 &= a(r,z) [\cos(\phi_0) - i \sin(\phi_0)]\\
E_\phi^1 &= a(r,z) [\sin(\phi_0) + i \cos(\phi_0)]\\
B_r^1 &= a(r,z) [-\sin(\phi_0) - i \cos(\phi_0)]\\
B_\phi^1 &= a(r,z) [\cos(\phi_0) - i \sin(\phi_0)].
\end{align}
Circularly and elliptically polarized lasers can obtained by adding two linearly polarized lasers with equal or non equal amplitudes and phase and polarization offset by $\pi/2$.

The time-evolution of electromagnetic fields are described by Faraday's and Ampere's equations (effectively written in normalized units),
\begin{align} 
   \frac{\partial \mathbf{B}}{\partial t} &= {- \nabla \times \mathbf{E}}, \\
   \frac{\partial \mathbf{E}}{\partial t} &= \nabla \times \mathbf{B} - \mathbf{J}.
\end{align}
Substituting the expansions for each field into these equations, we obtain the following equations for each mode, $m$, 
\begin{align}
\frac{\partial B_r^m}{\partial t} &= -\frac{i m}{r} E_z^m + \frac{\partial E_\phi^m}{\partial z} \label{maxwellbegin}\\
\frac{\partial B_\phi^m}{\partial t} &= - \frac{\partial E_r^m}{\partial z} + \frac{\partial E_z^m}{\partial r}\label{bphinonsingularity}\\
\frac{\partial B_z^m}{\partial t} &= - \frac{1}{r} \frac{\partial }{\partial r} (r E_\phi^m) + \frac{i m}{r} E_r^m \label{bzsingularity}\\
\frac{\partial E_r^m}{\partial t} &= \frac{i m}{r} B_z^m - \frac{\partial B_\phi^m}{\partial z} - J_r^m \label{ersingularity}\\
\frac{\partial E_\phi^m}{\partial t} &= \frac{\partial B_r^m}{\partial z} - \frac{\partial B_z^m}{\partial r} - J_\phi^m \\
\frac{\partial E_z^m}{\partial t} &= \frac{1}{r} \frac{\partial}{\partial r} (r B_\phi^m) - \frac{i m}{r} B_r^m - J_z^m \label{maxwellend}
\end{align}
We use different conventions for the coordinate system than that used  in reference, \cite{Lifschitz}, but the idea is identical. In vacuum, each mode evolves independently of every other mode. In addition, for a linear plasma response there is also no coupling between modes because under this limit each harmonic for the current  is only driven by the same harmonic for the fields. However, there is coupling between harmonics due to the macroparticle motion, i.e., nonlinear currents. The finite difference expression of these equations and associated complications near the r=0 axis will be discussed in Section~\ref{algorithm}.

\subsection{Symmetry properties of the axis \label{axissymmetry}}
When implementing the field equations expressed in cylindrical coordinates one inevitably comes across singularities at the axis ($r = 0$). The exact location of the singularities will depend on the layout of the grid values, but you can solve the singularities using symmetry-based arguments. As pointed out in ref \cite{Lifschitz}, G.S. Constantinescu et al.~\cite{polaraxis} discuss in detail how the field values behave at the cylindrical, $r = 0$, axis. To summarize, for any scalar and cartesian fields ($E_z,B_z$) only the $m = 0$ mode is  non-vanishing on the r=0 axis (this was already used in OSIRIS). On the other hand, for cylindrical field components ($E_r, B_r, E_\phi, B_\phi$), only the $m = 1$ mode is non-vanishing on the r=0 axis. 

\subsection{Boundary conditions for fields and particles}
Currently, we use conducting boundary conditions for the fields, and an absorbing boundary condition for the particle at $r = r^{\text{max}}$. We also use a moving window in the $z$ direction. In the future, more boundary conditions will be added in both $r$ and $z$, including the ability to launch a laser from a wall or a moving antenna.

\section{Algorithm}
\label{algorithm}

The truncated azimuthal mode geometry has been incorporated into the OSIRIS simulation framework. The electromagnetic fields were calculated on $2m + 1$ 2D grids; one grid representing the cylindrically symmetric (and real) $0^\text{th}$ order mode, while the rest represented the real and imaginary components of the higher order modes. Each field mode was advanced in accordance to Equations~\ref{maxwellbegin}-\ref{maxwellend}, whose implementation will be discussed in more detail in Section~\ref{fieldsolver}. The macroparticle values $(x,y,z,p_x,p_y,p_z)$ were stored in 3D coordinates. When the fields were interpolated onto the particles, the mode contributions were added together as per Equation~\ref{expansion}, and converted into Cartesian coordinates. The particles were then advanced according to the relativistic equations of motion,
\begin{align}
\frac{d}{dt}\mathbf{P} &= q (\mathbf{E} + (\mathbf{v}/c) \times \mathbf{B}),\\
\frac{d}{dt}\mathbf{x} &= (1/m\gamma) \mathbf{P},
\end{align}
where $q$ and $m$ are the macroparticle charge and mass, respectively. Using the motion of the particles the current can be deposited onto the $2m + 1$ 2D grids using a charge-conserving deposition scheme, particle by particle, as described in detail in Section~\ref{chconssection}.

\subsection{Field solver}
\label{fieldsolver}
\begin{figure}[h]
\centering
\includegraphics[width=0.7\textwidth]{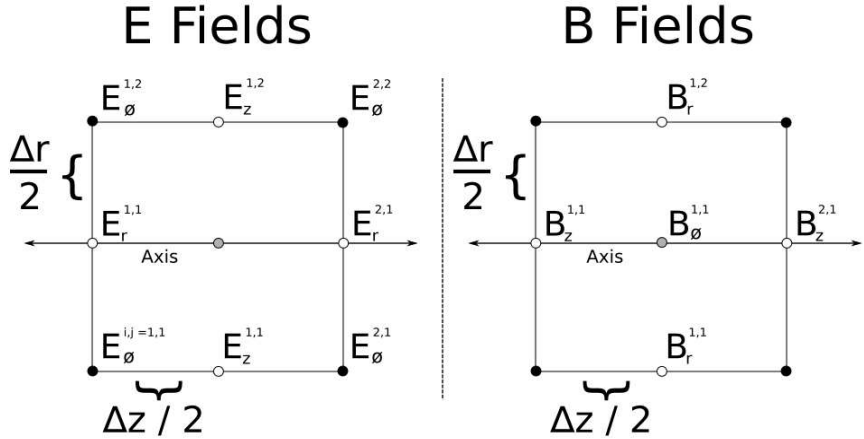}
\caption{The layout of the grid for the field components in relation to the cylindrical axis. The grid indices associated with the field point is indicated on the superscript. The $E_r, B_z$ and $B_\phi$ lie on the cylindrical axis for the axial cell.}
\label{staggering}
\end{figure}

The Maxwell's Equations~\ref{maxwellbegin}-\ref{maxwellend} for each harmonic were discretized over a uniform grid defined on the Yee Lattice \cite{yee}. Due to staggering, fields of the same index lie in different positions with respect to the axis, as is shown in Figure~\ref{staggering}. Some field values reside exactly on the cylindrical axis, which in this case will cause a singularity when solving Equations~\ref{bzsingularity} and \ref{ersingularity}.  It is important to note here that the location of the axis in our simulation is different from that of Lifschitz~\cite{Lifschitz}, where the equations for $B_r$ and $E_z$ present a singularity. Although the singularities occur for different field components, the logic with which we resolve these issues are effectively the same.

As discussed in Section~\ref{axissymmetry}, the axial fields are usually zero. The only axial fields we need to solve for are $B_z^0$, $E_r^1$, and $B_\phi^1$; the last of which does not pose a singularity. We use the integral form of Faraday's Law to find $B_z^0$ on the axis by integrating $\oint \mathbf{E} \cdot d\mathbf{l}$ in a loop around the axis, resulting in
\begin{equation}
\label{bzaxis}
B_z^{0, i,1} = B^{0,i,1}_z - 4 \frac{\Delta t}{2} \times \frac{E^{0,i,1}_\phi}{\Delta r}.
\end{equation}
This is the same method that was used in the 2D Cylindrical simulation that was already implemented in OSIRIS. As for the $m = 1$ mode, we use the fact that
$$ \lim_{r \rightarrow 0} \frac{im}{r} B_z^m =  i m \frac{\partial B_z^m}{\partial r} ,$$
to define 
\begin{align}
\frac{im}{r}B_z^{m,i,1} &= i m \left. \frac{\partial B_z}{\partial r}\right|^{r = 0} = i m \frac{B_z^{m,i,2}}{\Delta r}, \nonumber 
\end{align}
where the fact that $B_z^{m,i,1} = 0$ was used. We then obtain a nonsingular expression for $\frac{\partial E_r^1}{\partial t}$ at the axis,
\begin{equation}
\label{eraxis}
\frac{\partial E_r^{m=1,i,j=1}}{\partial t} =  i \frac{1}{\Delta r} B_z^{1,i,2} - \frac{1}{\Delta z}\left(B_\phi^{1,i,1} - B_\phi^{1,i-1,1}\right) - J_r^1,
\end{equation}
where $j = 1$ is the radial index of the cell that sits on the axis.

\subsection{Charge Conserving Current Deposition}
\label{chconssection}
In OSIRIS Gauss's law is maintained by using a rigorously charge conserving deposit. For example, in 3D Cartesian geometry the particles have a shape 
$$ S_x (x - x_p(t)) S_y (y - y_p(t)) S_z(z - z_p(t)),$$
and the cell corners (where the charge density is defined) are defined at $x_g, y_g, z_g$.  The charge density at a time $t$ at the grid locations is therefore,
$$ \rho = S_x(x_g - x_p(t)) S_y(y_g - y_p(t)) S_z(z_g - z_p(t)).$$
The current is defined at different locations (the cell faces) and is defined such that
$$ \frac{\overline{\partial} }{\partial t} \rho + \overline{\nabla} \cdot \mathbf{J} = 0,$$
where $\overline{~}$ indicates a finite difference representation of the derivative, e.g.,
\begin{align}
\left. \frac{\overline{\partial}}{\partial t} \rho \right|^{t + \Delta t/2} = \frac{1}{\Delta t} &\left[ S_x (x_g - x_p(t + \Delta t)) S_y (y_g - y_p(t + \Delta t)) S_z (z_g - z_p(t + \Delta t))\right. \nonumber\\
&~~~~~~~~~~~~~~~~~~~~~- \left.S_x (x_g - x_p(t )) S_y (y_g - y_p(t )) S_z (z_g - z_p(t ))\right].
\end{align}

There is not a unique solution for a $\mathbf{J}$ such that $\frac{\overline{\partial}}{\partial t} \rho + \overline{\nabla} \cdot \mathbf{J} = 0$ as one can always add a curl to one solution. To determine $\mathbf{J}$, OSIRIS implements the Density Decomposition method described by Esirkepov \cite{Esirkepov01}, which is the generalization of the method developed by Villasenor and Buneman \cite{chargecons} for arbitrary particle shapes. In the method of Villasenor and Buneman, linear particle shapes are assumed.
If the particle motion stays within a cell then and moves from $x_i, y_i$ to $x_f,y_f$ in one time step, this method can be viewed as averaging the current contribution over all paths that are decomposed into segments that include motion orthogonal to a cell face.  
If the particle motion crosses cell boundaries, then the motion is split into segments lying entirely inside individual cells, and the method is applied to each individual segment. 

Extending the charge conserving current deposit to 2D $r$-$z$ is relatively straight forward because the cells are still rectangular. In this case, one needs to recognize that $S_r (r - r_p(t))$ includes a $1/r$ term. Viewed another way each simulation particle represents a fixed amount of charge so as it moves closer to the $r = 0$ axis the charge density must increase. The $J_\phi$ component is simple to define in such a code as it is simply $\rho V_{\phi,p}$ defined on the r-z grid. 

On the other hand, it is not straightforward to define $J_\phi$ in the new algorithm. In particular, for the $m = 0$ harmonic the standard method works but for the $m \neq 0$ harmonics more thought is needed. However, as we show next, $J_\phi^m$ can be determined using the $J_{\perp}$ from  existing charge conserving deposition scheme for the m=0 mode. We begin from the definition of the particle shape in cylindrical coordinates.
$$ S \equiv S_r (r - r_p(t)) S_\phi (\phi - \phi_p(t)) S_z(z - z_p(t))$$
so the charge density is $Q S$. We also note that $S_r$ has a $1/r$ factor so that
$$ \int dr r d\phi dz S = 1.$$
The particle positions are known at full integer values of time, $t +  \Delta t n$, and the particle momentum (and velocity) are known at half integer values of time, $t + \Delta t (n + \frac{1}{2})$. In addition, the currents are only defined on the $r$-$z$ grid, i.e.,  there is no grid in $\phi$.

Next, we look for solutions for $\mathbf{J}$ that satisfy the finite difference operator version of the continuity equation
\begin{align}
\left. \frac{\overline{\partial}}{\partial t} \rho \right|^{n + \frac{1}{2}} &= \sum_p q \left[ S_r (r - r_p^{n + 1}) S_\phi (\phi - \phi_p^{n + 1}) S_z (z - z_p^{n + 1})\right. \nonumber\\
&~~~~~~~~~~~~~~~~~~~~-  \left.S_r (r - r_p^{n }) S_\phi (\phi - \phi_p^{n }) S_z (z - z_p^{n }) \right] \nonumber \\
&= - \overline{\nabla} \cdot \mathbf{J}^{n + \frac{1}{2}}.
\end{align}
The next step is to expand $S_\phi$ in global basis functions (azimuthal harmonics),
\begin{equation}
\label{shapephi}
 S_\phi (\phi - \phi_p) = \sum_m S_{\phi,m}(\phi_p) e^{i m \phi}, 
\end{equation}
where
$$ S_{\phi,m} \equiv \int_0^{2 \pi} \frac{d \phi^\prime}{2 \pi} e^{- i m \phi} S_\phi (\phi^\prime - \phi_p).$$
If $S_\phi \equiv \delta(\phi - \phi_p)$ then $S_{\phi,m} = \frac{1}{2\pi} e^{- i m \phi_p}$. In addition, $\rho$ and $\mathbf{J}$ defined on the $r$-$z$ grid are expanded in azimuthal harmonics

\begin{equation}
\label{rhoexpand}
\begin{pmatrix}
      \rho    \\
      \mathbf{J}  
\end{pmatrix} 
= \sum_m
\begin{pmatrix}
     \overline{\rho}_m (r_g,z_g,t)   \\
      \overline{\mathbf{J}}_m (r_g, z_g,t) 
\end{pmatrix} e^{i m \phi},
\end{equation}

where the $\overline{~}$ refers to a quantity defined only on the grid. The continuity equation can be written as
\begin{equation}
\label{conteq1}
 \frac{\overline{\partial}}{\partial t} \rho + \overline{\nabla}_\perp \cdot \mathbf{J}_\perp + \frac{1}{r} \frac{\partial}{\partial \phi} J_\phi  = 0, 
\end{equation}
where $\perp$ refers to the $r$-$z$ plane. Substituting Equations~\ref{shapephi} and \ref{rhoexpand} into Equation~\ref{conteq1} gives
\begin{align}
\sum_m e^{i m \phi} &\left\{ \sum_p \frac{1}{\Delta t} \left[ S_r(r_g - r_p^{n + 1}) S_{\phi,m}(\phi_p^{n + 1}) S_z(z - z_p^{n + 1})\right.\right. \nonumber \\
& ~~~~~~~~~~~~~~~~~~~- \left.  S_r(r_g - r_p^{n}) S_{\phi,m}(\phi_p^{n}) S_z(z - z_p^{n}) \right] \nonumber \\
&~~~+ \left.  \overline{\nabla}_\perp \cdot \overline{\mathbf{J}}_{\perp,m}^{n + \frac{1}{2}}  + \frac{i m}{r} \overline{J}_{\phi,m}^{n + \frac{1}{2}} \right\} = 0 \label{sumsummode}
\end{align}
We next recognize that by definition, for each particle
\begin{align}
\overline{\rho} &= \sum_m \overline{\rho}_m e^{i m \phi}, ~~\text{and} &\overline{\rho}_m =  \overline{\rho}_0 S_{\phi,m},
\end{align}
where $\overline{\rho}_0$ is the charge for one particle on the $r$-$z$ grid for the $m = 0$ mode (recall $S_{\phi,0} = 1$ by normalization). Likewise
\begin{align}
\overline{\mathbf{J}}_\perp &= \sum_m \overline{\mathbf{J}}_{\perp,m} e^{i m \phi}, ~~\text{and} &\overline{\mathbf{J}}_{\perp,m} =  \overline{\mathbf{J}}_{\perp,0} S_{\phi,m}
\end{align}
in the continuous time limit. We next show that using these definitions and a $\overline{\mathbf{J}}_{\perp,0}$ defined to conserve charge for the m=0 mode (what already existing in OSIRIS)  leads to an expression for $\overline{J}_\phi$. Substituting these expressions into Equation~\ref{sumsummode} gives for each $m$ and $p$ in the sum
\begin{align}
\frac{1}{\Delta t} &\left[ S_r (r_g - r_p^{n + 1}) S_{\phi,m}(\phi_p^{n + 1}) S_z(z_g - z_p^{n + 1}) - S_r (r_g - r_p^{n }) S_{\phi,m}(\phi_p^{n }) S_z(z_g - z_p^{n })\right] \nonumber \\
& ~~~~~~~~~~~~~~~~~~~~~+ S_{\phi,m} (\phi_p^{n + \frac{1}{2}}) \overline{\nabla}_\perp \cdot \overline{\mathbf{J}}_{\perp,0}^{n + \frac{1}{2}} + \frac{i m}{r} \overline{\mathbf{J}}_{\phi,m}^{n + \frac{1}{2}} = 0.
\end{align}
By definition $\frac{\overline{\partial}}{\partial t} \overline{\rho}_0 + \overline{\nabla}_\perp \cdot \overline{\mathbf{J}}_{\perp,0} = 0$, so we are left with
\begin{align}
\overline{J}_{\phi,m}^{n + \frac{1}{2}} = i \frac{1}{\Delta t} \frac{r}{m} &\left\{ S_r (r_g - r_p^{n + 1}) S_z(z_g - z_p^{n + 1}) \left[ S_{\phi,m}(\phi_p^{n+1}) - S_{\phi,m}(\phi_p^{n + \frac{1}{2}})\right]\right. \nonumber \\
&~~~~~~~~~~- \left. S_r (r_g - r_p^{n}) S_z(z_g - z_p^{n}) \left[ S_{\phi,m}(\phi_p^{n}) - S_{\phi,m}(\phi_p^{n + \frac{1}{2}})\right]\right\},
\end{align}
For $S_\phi = \delta(\phi - \phi_p)$ we have $S_{\phi,m} = \frac{1}{2\pi}e^{- i m \phi_p}$. We then  define $\overline{\phi}_p \equiv \frac{\phi_p^{n + 1} + \phi_p^n}{2}$ and $\Delta \phi_p \equiv \phi_p^{n + 1} - \phi_p^n$, to obtain the result we use in OSIRIS for each particle,
\begin{align}
\overline{J}_{\phi,m} = i \frac{r}{m} \frac{1}{\Delta t} \frac{e^{- i m \overline{\phi}_p}}{2 \pi}&\left[ S_r (r_g - r_p^{n + 1}) S_z(z_g - z_p^{n+1}) \left( e^{i m \frac{\Delta \phi}{2}} - 1\right)\right. \nonumber \\
&~~~~~ - \left. S_r (r_g - r_p^{n }) S_z(z_g - z_p^{n}) \left( e^{- i m \frac{\Delta \phi}{2}} - 1\right)\right] ,\label{jphichargecons}
\end{align}
where the particle shapes in $r$ and $z$ are still general. Currently, OSIRIS implements linear, and quadratic, interpolation for the new algorithm.

\subsection{Complex Exponentials}
When evaluating expressions like Equation~\ref{expansion} and  \ref{jphichargecons}, you need to evaluate the complex exponential $e^{i m \phi}$. The particle variables are stored in Cartesian coordinates, and nowhere in the simulation is $\phi$ directly stored or calculated. In addition, evaluating trigonometric functions will be computationally inefficient. Instead, we use double and triple angle formulas to obtain these values up to $m = 4$,
\begin{align}
e^{i \phi}~ &= ( x + i y ) / r \\
e^{i 2 \phi} &= ((x^2 - y^2) + 2 i x y) / r \\
e^{i 3 \phi} &= (4 x^3/ r^3 - 3x/r) - i (y^3/r^3 - 3y/r)\\
e^{i 4 \phi} &= ((x^2 - y^2)^2  - 4x^2 y^2)/r^4 + 4 i x y (x^2 - y^2) / r^4.
\end{align}
This same optimization is used by Lifschitz \cite{Lifschitz}. In order to calculate $e^{- i m \phi}$ one only needs to swap the sign of the imaginary part. One may extend this method to an arbitrary number of modes using the exponential relation $e^{i m \phi} = e^{i \phi} \times e^{i (m - 1) \phi}$, which is what is done in OSIRIS to capture any number of modes specified by the user.

\section{Results}
\label{results}

In this section we present examples from simulations using the new algorithm. We present simulation results for a laser wakefield accelerator (LWFA), a plasma wakefield accelerator (PWFA), and an LWFA case with beam loading (combining the laser and beam propagation capabilities) case respectively. In \cite{Lifschitz} only an LWFA example was given. We also demonstrate the degree to which Gauss' law is conserved with the new current deposit algorithm. The new algorithm has many more potential applications than LWFA and we will discuss some in the conclusions and directions for future work section. For the ``hybrid'' r-z simulations, we typically use 2 particles per cell in the $r$-$z$ directions, and $8$ or $16$ particles distributed evenly over $0 \leq \phi < 2 \pi$ (The particles are distributed along spokes at each $z$). The former can be considered as 16 particles per cell when comparing to the speed up from the full 3D simulation. The effect of the particle resolution in $\phi$ will be discussed for some of the examples, but for these simulations $8$ particles appeared to be enough to capture the physics. We note that different methods for initiating the particles can be considered and we leave this for future work.

For the LWFA simulations, we model the example given in Lu et al., \cite{weilu}. In this example a circularly polarized $200TW$, $30fs$, $0.8 \mu m$ laser pulse with a spot size of $19.5 \mu m$ propagates through  a fully ionized plasma of density $n = 1.5 \times 10^{18}cm^{-3}$. The  laser has a normalized vector potential of magnitude $a_0 = 4$. In ref. \cite{weilu}, it was found using full 3D OSIRIS simulations that such a laser could generate an ultrashort ($10fs$) self-injected mono-energetic bunch with an energy centered at $1.5 GeV$. We have reproduced the 3D simulation for this paper using quadratic splines (linear splines were used in \cite {weilu}). 
\subsection{Charge Conservation Tests}
\label{chconstests}
\begin{figure}[htbp] 
   \centering
   \includegraphics[width=0.6\textwidth]{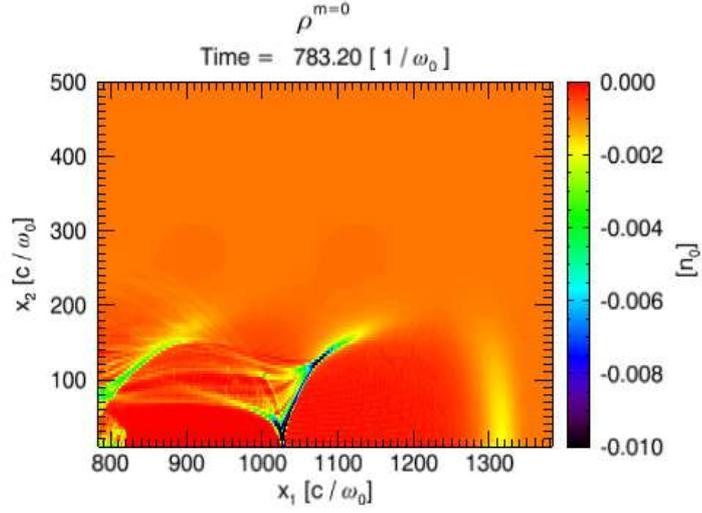} 
   \caption{A wake formed by a circularly polarized laser penetrating $0.1mm$ into the plasma, presented here in the $m = 0$ azimuthal mode of the charge density. The charge conservation tests presented in Figure~\ref{chcons} correspond to this result of this simulation. If you take the divergence of the electric fields in mode $0$, it will correspond to this plot exactly. }
   \label{rho1}
\end{figure}

\begin{figure}[htbp] 
   \centering
   \includegraphics[width=0.6\textwidth]{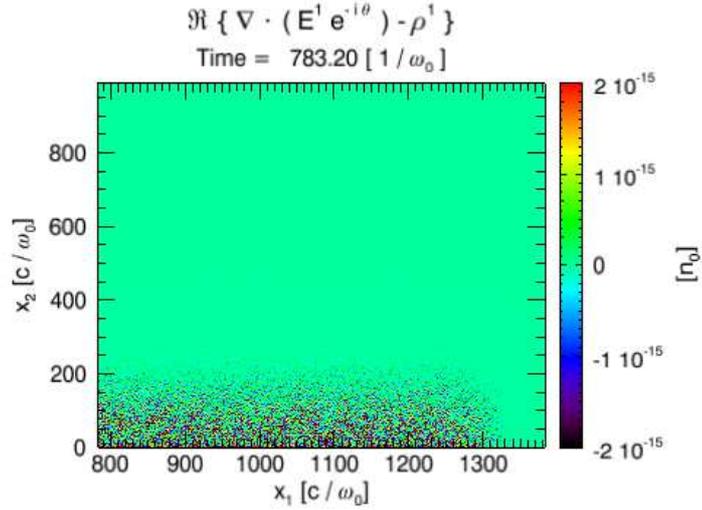} 
   \caption{The deviation of the charge conservation (Gauss' law) of the real part of mode $m = 1$, for a simulation utilizing quadratic particle interpolation. The deviation in the Gauss' law is maintained to within the accuracy of double precision arithmetic. }
   \label{chcons1}
\end{figure}

We start by checking the degree to which charge, i.e., Gauss' law is conserved.  To test the effectiveness of the deposition scheme in Equation~\ref{jphichargecons}, we need to carefully examine Gauss' law for the new algorithm. 
We begin by expanding Gauss' law as per Equation~\ref{expansion},
\begin{align}
&\nabla \cdot \mathbf{E} - \rho = \Re\left\{ \sum_m \nabla \cdot\left[ \mathbf{E}^m(r,z) e^{ i m \phi}\right] - \sum_m\rho^m(r,z) e^{ i m \phi}\right\} = 0,\\
\implies  &\nabla_\perp \cdot \mathbf{E}^m + i \frac{m}{r}E_\phi - \rho^m = \frac{1}{r}\frac{\partial}{\partial r}\left( r E_r\right) + \frac{\partial E_z}{\partial z} + i \frac{m}{r} E_\phi - \rho^m= 0. \label{chconsdiag}
\end{align}
This means that the charge of each mode must be conserved independently from each other mode, and that the divergence is simultaneously affected by both the real and imaginary parts of the fields. We used the LWFA case described above as the test case. For these tests we used a smaller computational window of dimensions $76.4 \mu m \times 127 \mu m$, and $3000 \times 256$ grid points. We let the laser pulse propagate ($0.1mm$)  into the plasma, so that a well defined wake is formed as shown in Figure~\ref{rho1}.  We used $2$ particles per cell in the $r-z$ direction and $8$ particles in the $\phi$ direction, giving a total of $16$ particles per cell. We ran simulations keeping up to the $2$nd harmonic and the charge conservation of each mode was tested rigorously. Both linear and quadratic interpolations were tested (the particle shape in the $\phi$ direction was a delta function). The output of the charge conservation diagnostic for the real part of mode $1$ (for which the numerical noise was the greatest) is shown in Figure~\ref{chcons1}. In these simulations we used double precision floating point numbers, which have $15$ numerical orders of accuracy in decimal units. When subtracting two nearly identical numbers, a roundoff error $10^{-15}$ below the working order of magnitude, which in this case is $2.83$ will be observed.  
In Figure~\ref{chcons1} charge is conserved to within the roundoff error of the double precision arithmetic at each grid point of the simulation. The noise is slightly larger as $r$ approaches zero, since the field values are scaled to $1/r$ when calculating the divergence.

\begin{figure}[htbp] 
\centering
\begin{tabular}{cc}
\includegraphics[width=0.45\textwidth]{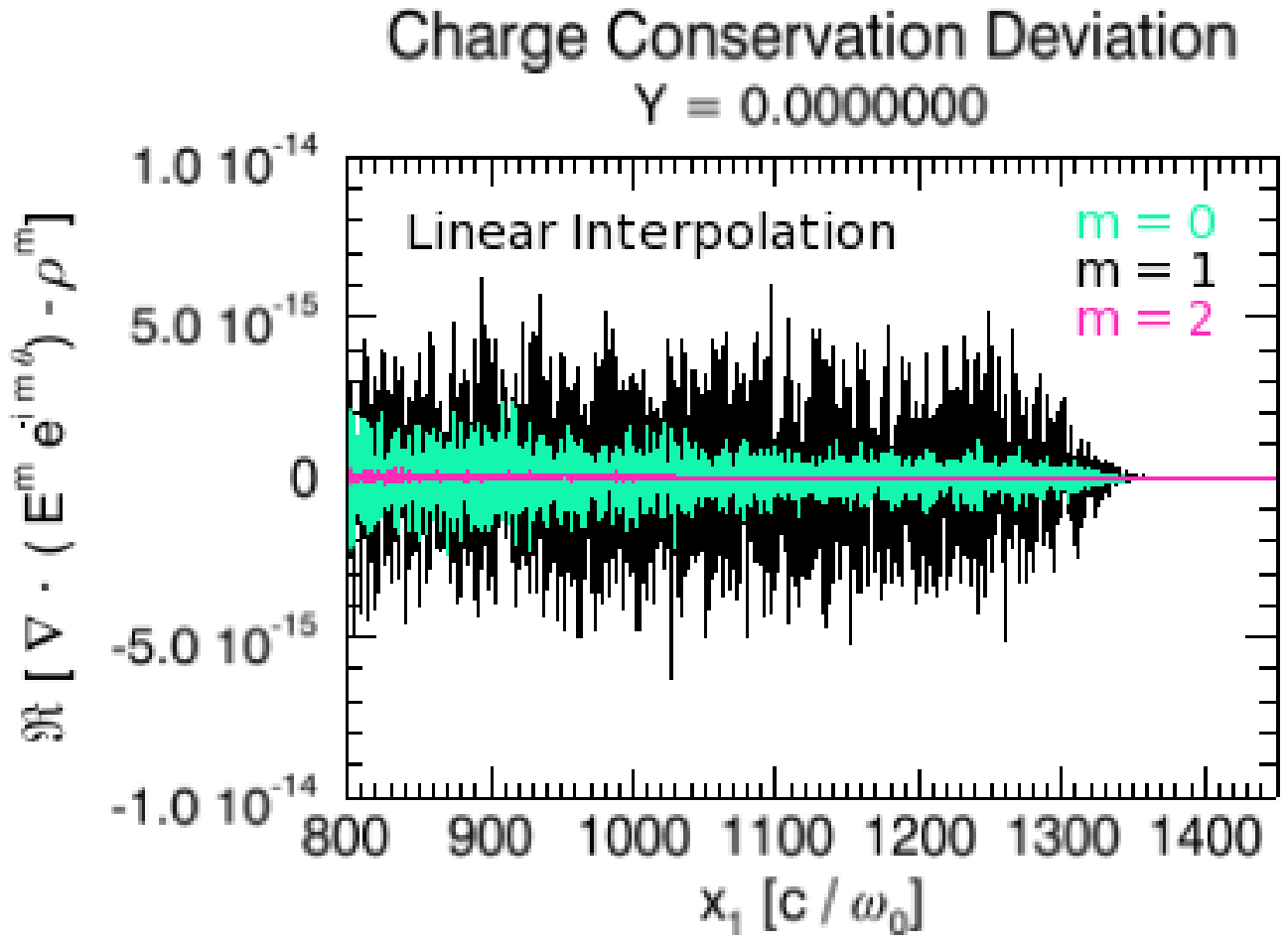} 
   \includegraphics[width=0.45\textwidth]{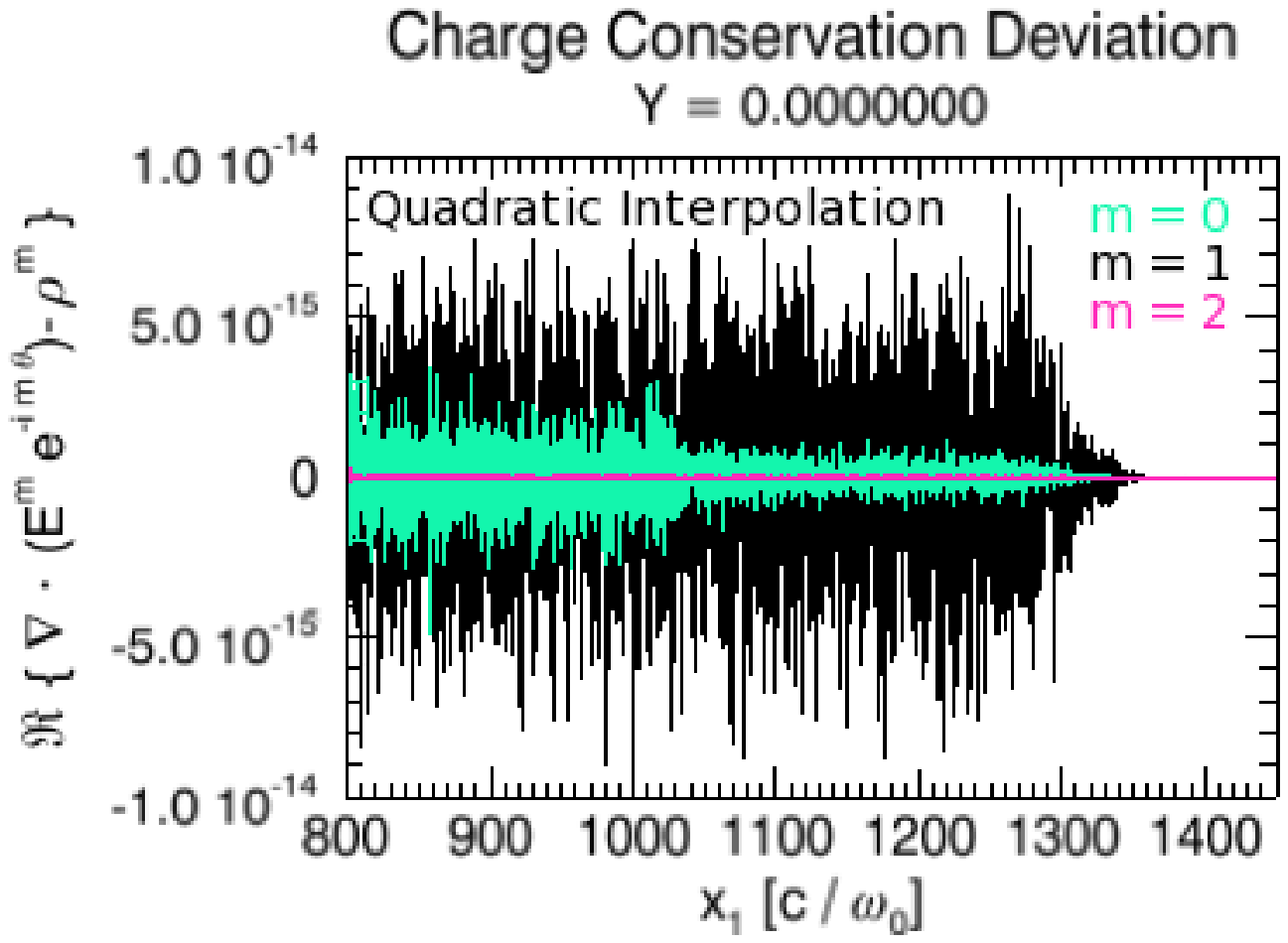} \\
 \end{tabular}
   \caption{The deviation of charge conservation (Gauss' law), along $r=0$, for modes $0$, $1$, and $2$. Quadratic interpolation (right) shows a slightly larger roundoff error then the linear interpolation result (left). The error in $m = 1$ (the component with the laser) is the largest in both cases.}
   \label{chcons}
\end{figure}

The lineout along the axis of the charge conservation for each mode  is presented in Figure~\ref{chcons}. The numerical noise was slightly higher for the quadratic interpolation than the linear interpolation, but in each case the charge conservation was satisfied to roundoff error for every mode. In addition, the largest residual error is in the m=0 harmonic which includes the laser field. These tests validate the use of Equation~\ref{jphichargecons} and the existing current depositing algorithm for the m=0 harmonic in the $r$-$z$ grid. We have also tested the charge conservation for many more cases. 

\subsection{Comparison of LWFA Results with 3D Simulations}

\begin{figure}[htbp] 
\centering
\begin{tabular}{cc}
 \includegraphics[width=0.5\textwidth]{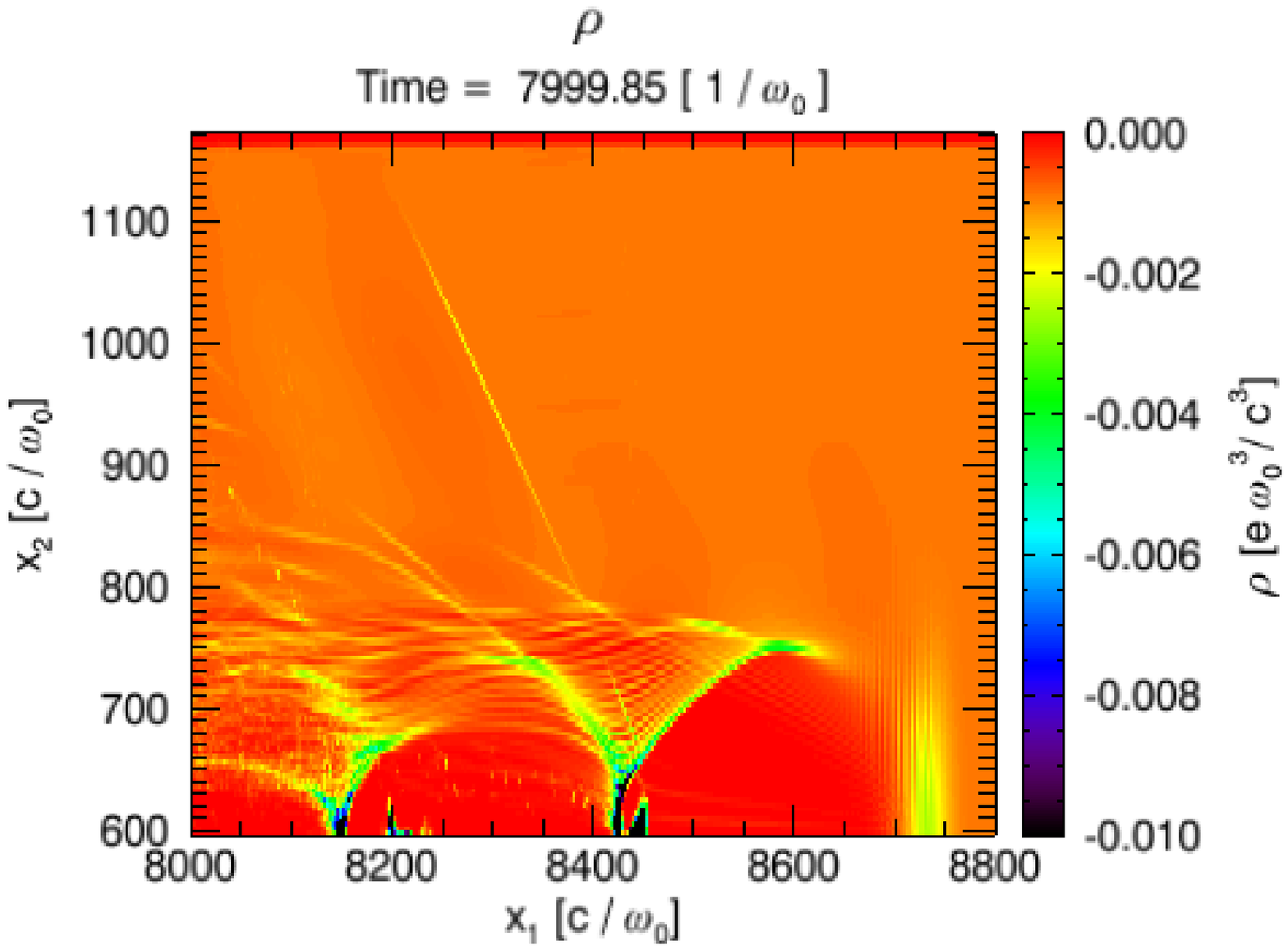} 
      \includegraphics[width=0.5\textwidth]{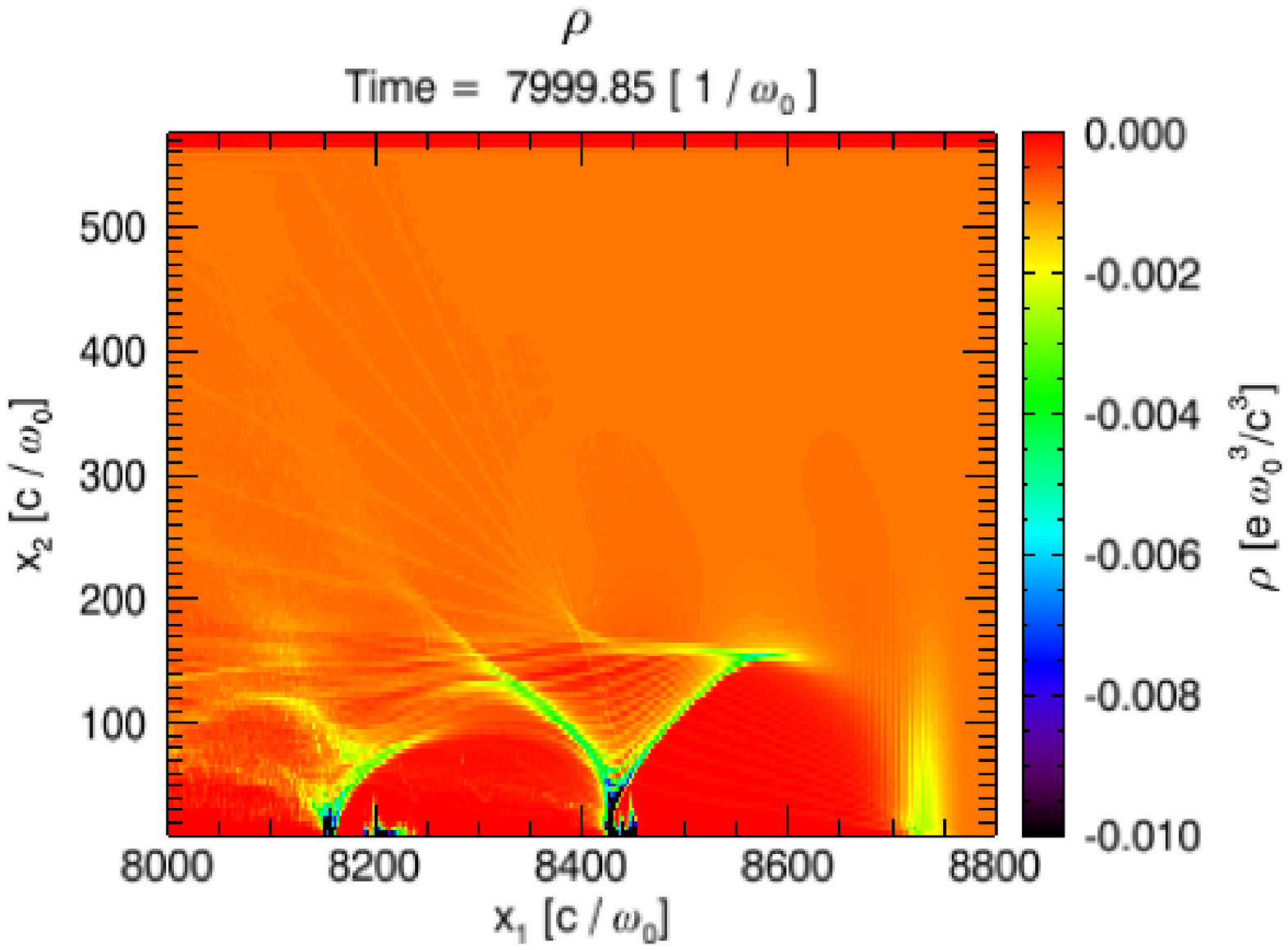} \\
\end{tabular}
      \caption{These are the electron charge density distributions for the full 3D simulation at 0.1cm (left) and the cylindrical mode simulation at 0.1cm (right). The cylindrical mode density cross-sections are taken at $\phi = 0$, which corresponds to the top half of the cross section of the 3D simulation at $y = 0.0$. Both simulations used quadratic interpolation, and both simulations used the same cell sizes. }
   \label{dens}
\end{figure}

We next present results from the LWFA simulation described earlier  keeping only up to the $m = 1$ harmonic. The LWFA simulation discussed at the beginning of this section was run to about $0.1cm$. For the full 3D simulation,  a $4000\times300\times300$ grid with dimensions $101.9 \mu m \times 149.2 \mu m \times149.2 \mu m$ was used with $2$ particles per cell . The time step was chosen as close as possible to the Courant limit. The hybrid r-z simulation used a computational window of dimensions $101.9 \mu m \times 74.6 \mu m$ and $4000 \times 150$ grid points. The simulation was conducted with $2$ particles per cell in the $r$-$z$ directions with $16$ particles in the $\phi$ direction.  For typical LWFA simulations, the $m=1$ mode captures enough modal asymmetry to effectively simulate the physics for round laser beams without any tilts. In later publications we will describe the additional physics that can be studied by including more harmonics. Note that the wake excited by a linearly or circularly polarized  cylindrically symmetric laser is itself cylindrically symmetric. For the hybrid r-z simulations we use a time step as close to the stability limit as possible. We note that we empirically found that this limit is close to the 3D Courant limit where we use an "effective" cell size in the $\phi$ direction roughly given by $\Delta r  \pi/m_{max}$ where $m_{max}$ is the highest harmonic kept. 
In addition, we found that only $8$ particles across $0 \leq \phi < 2 \pi$ were needed to avoid substantial noise in the first bubble. The signal-to-noise ratio scales as $\sqrt{m_{max}}$, so the fewer modes you use the fewer particle resolution you need across the $\phi$ coordinate\cite{Lifschitz}. Therefore, the effective speed up is roughly proportional to the number of simulation particles. In a 3D simulation $n_{p}^{3D} = N_x N_y N_z N_{pc}$ particles are uses, where $N_x, N_y,$ and $N_z$ are the number of cells in each Cartesian directions, and $N_{pc}$ is the number of particles per cell.  In the 2D hybrid simulation it is $n_p^{\text{2D-hybrid}} = N_z \frac{N_x}{2} N_{p, \text{$r$-$z$}} N_{p,\phi}$, where $N_{p, \text{$r$-$z$}}$ is the number of particles in the $r$-$z$ plane and $N_{p,\phi}$ is the number of particles distributed over $0 \leq \phi < 2 \pi$.

\begin{figure}[htbp] 
    \centering
    \begin{tabular}{cc}
    \includegraphics[width=0.45\textwidth]{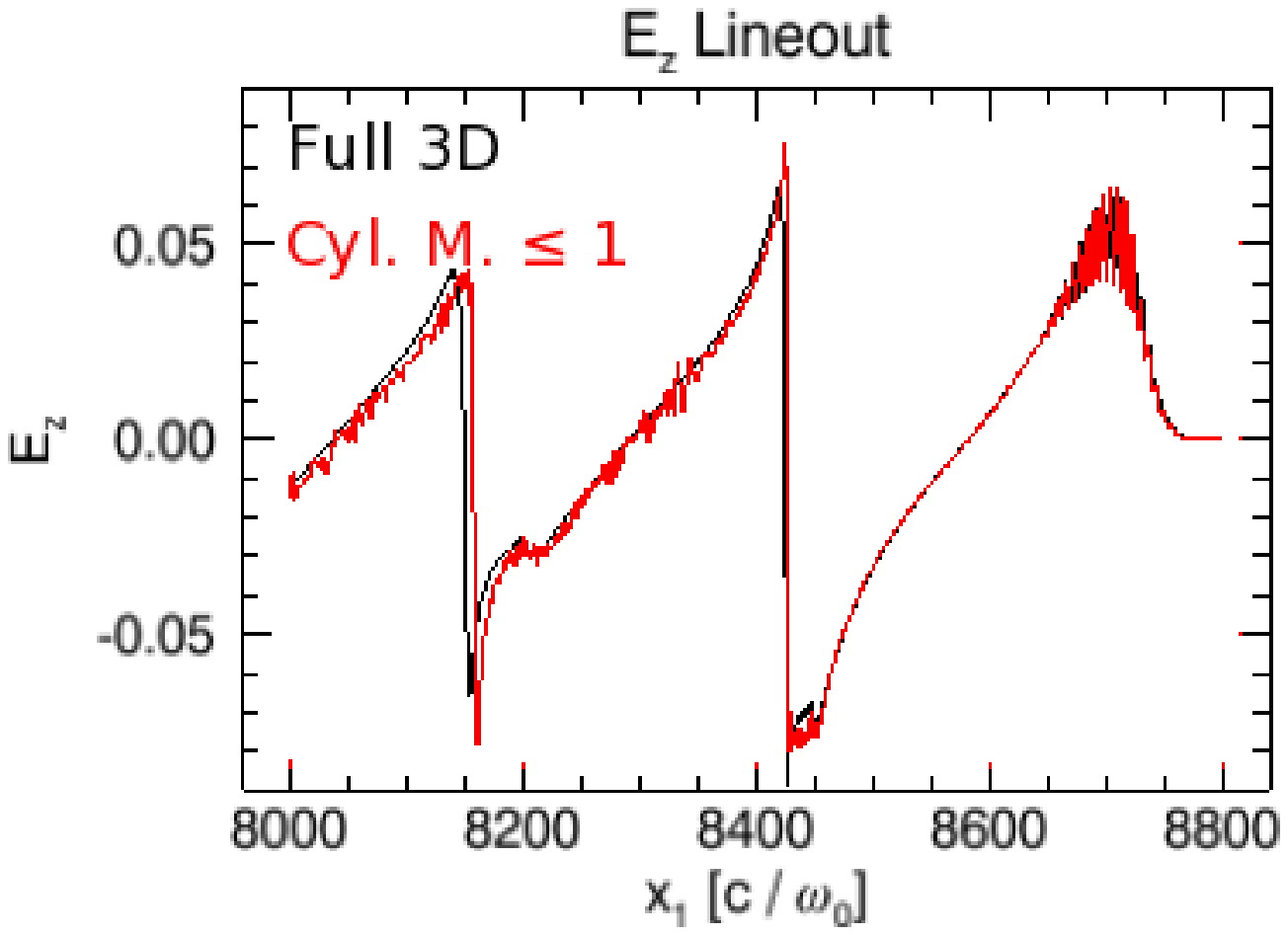} 
    \includegraphics[width=0.45\textwidth]{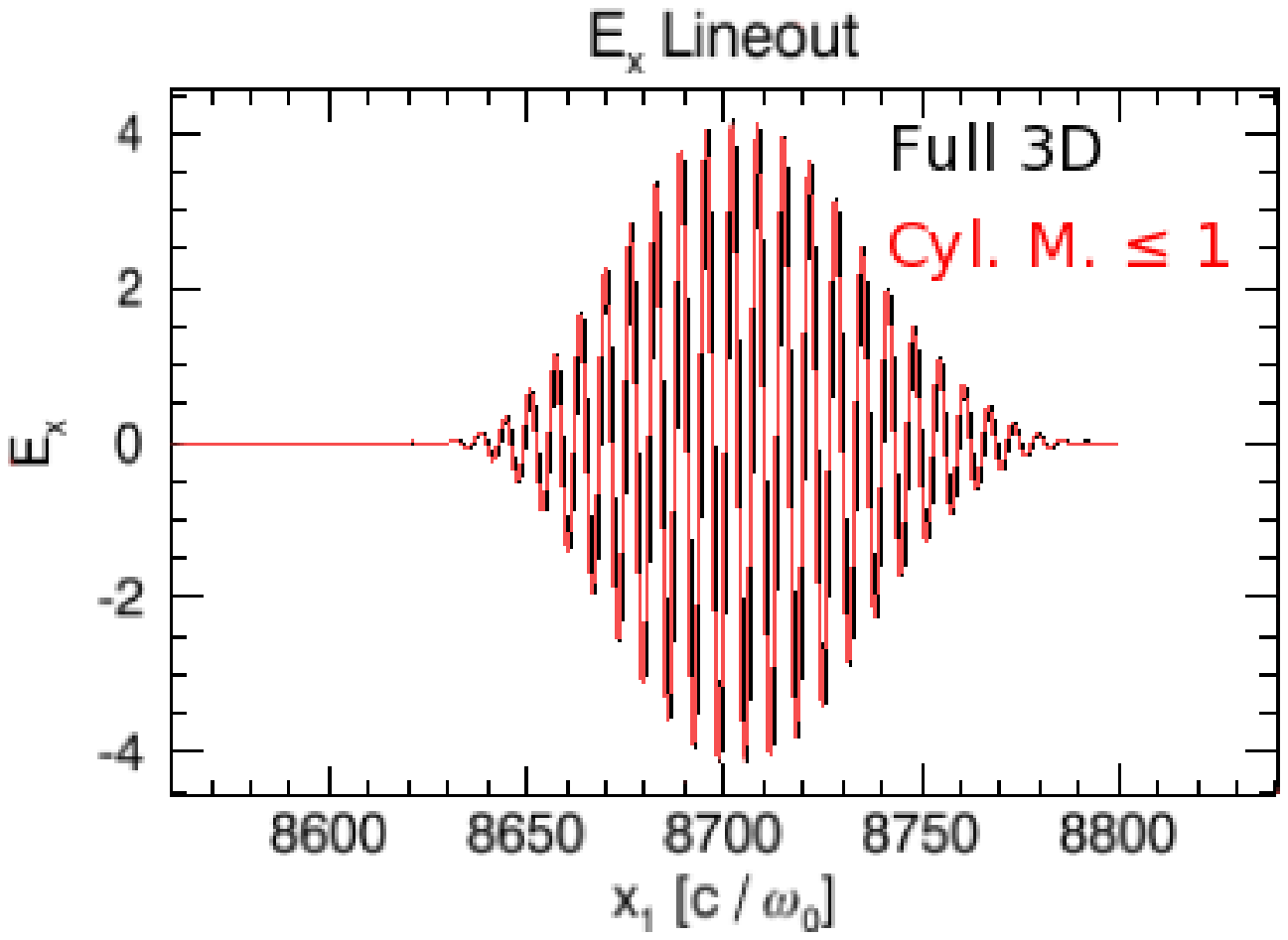} \\
    \end{tabular}
    \caption{Lineouts along the laser for the $E_z$ and $E_x$ fields for the 3D (black) and 2D hybrid (red) simulations. The lineout of $E_x$ is zoomed in to more easily see the matching of the phase of the laser.}
    \label{elineouts}
 \end{figure}

\begin{figure}[htbp] 
   \centering
   \includegraphics[width=0.5\textwidth]{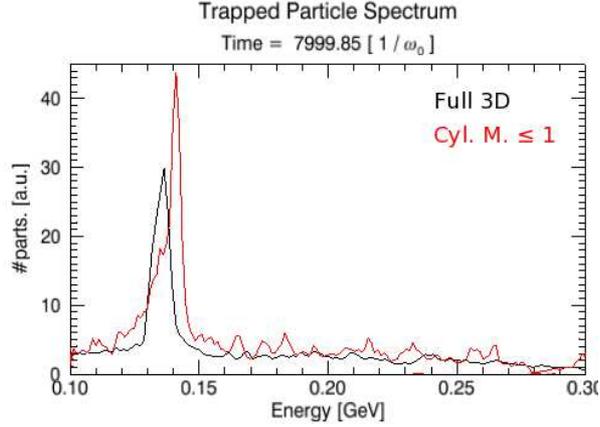} 
   \caption{The spectrum of the trapped particles from the 3D (black) and 2D hybrid (red) simulations. The laser has propagated $0.1cm$ into the plasma.}
   \label{spectrums-combined}
\end{figure}

\begin{figure}[htbp] 
   \centering
   \begin{tabular}{cc}
   \includegraphics[width=0.5\textwidth]{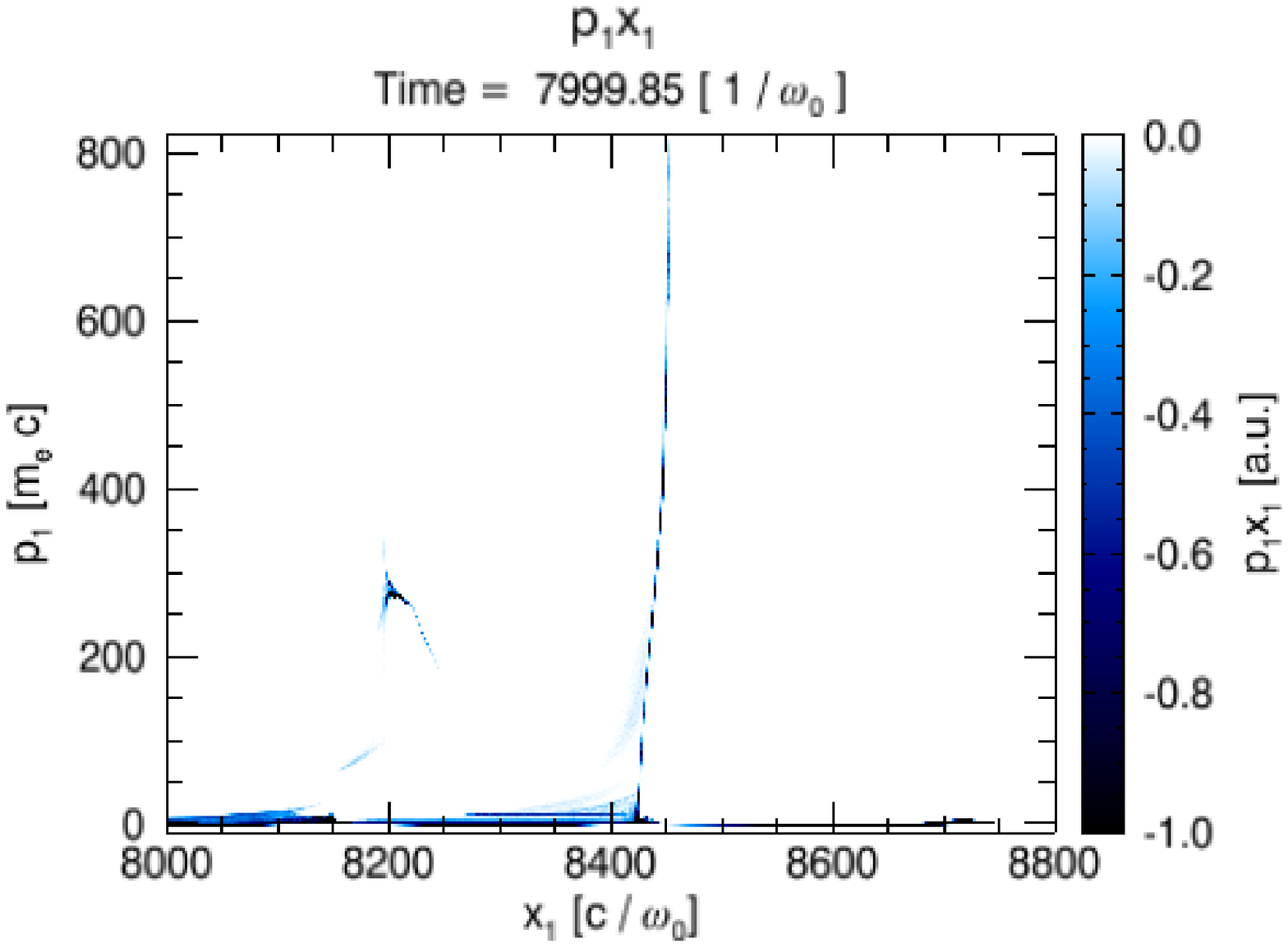} 
   \includegraphics[width=0.5\textwidth]{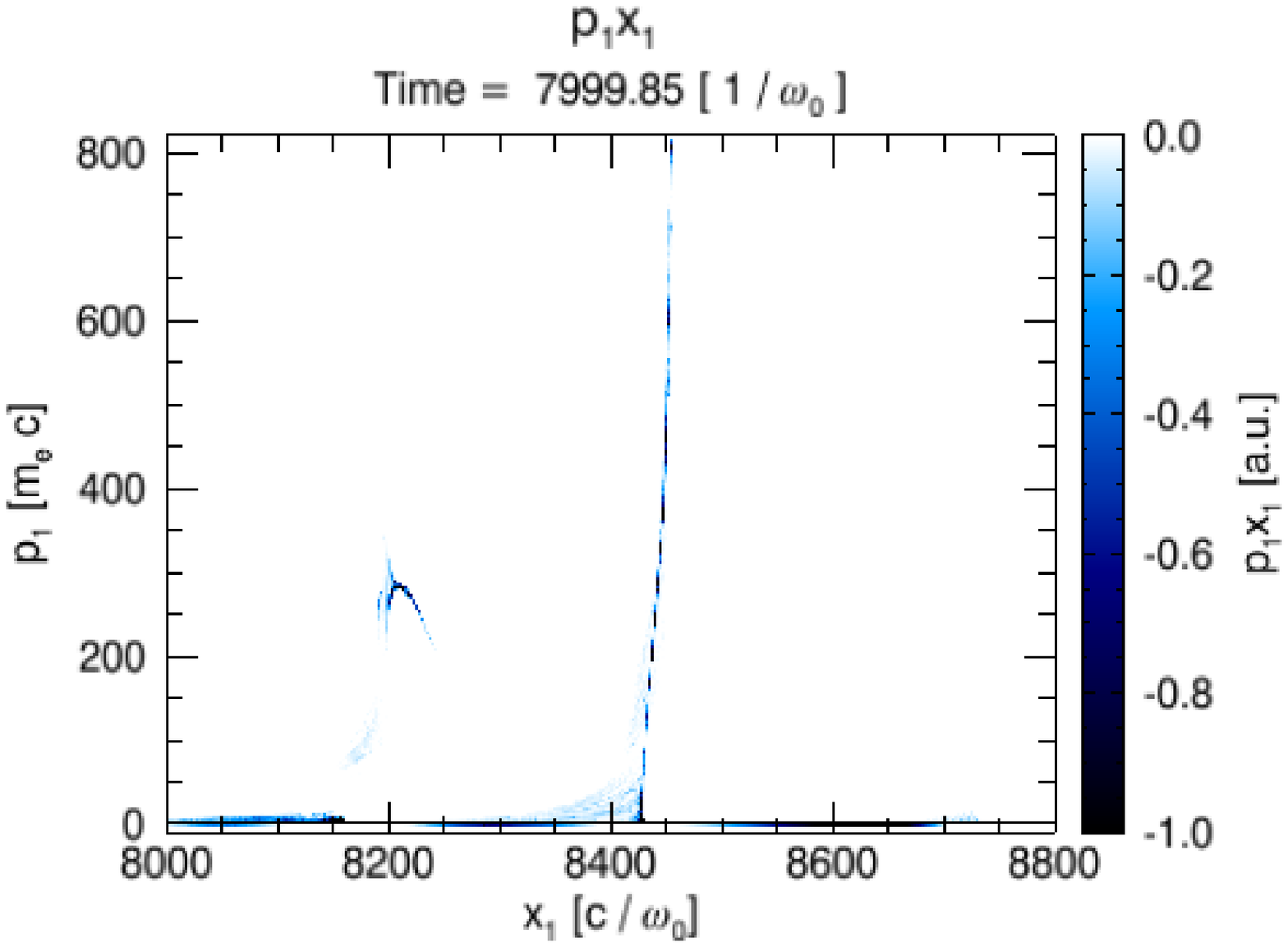} \\
   \end{tabular}
   \caption{The $p_1 x_1$ phase-space plot for the 3D (left) and 2D hybrid (right) simulations, after the laser had propagated $0.1cm$ into the plasma. }
   \label{p1x1plots}
\end{figure}

   
We show results after the  laser was propagated through the plasma over a distance of $0.1 cm$. Two-dimensional density plots corresponding to a cut across the data of the 3D simulation or the $\phi=0$ plane for the hybrid simulation are shown in Figure~\ref{dens}. 
The 2D density contours for the wake were identical throughout most of the simulation, aside from a small number of particles which had been trapped late in the full 3D simulation. However, this did not significantly affect the acceleration process of the mono-energetic bunch. The accelerating electric field of the 3D and the 2D hybrid modal simulations are  shown on the left hand side, and the laser profiles are shown on the right hand side of Figure~\ref{elineouts}.
The spectrum of the trapped particles in the two cases are shown in Figures~\ref{spectrums-combined} and  longitudinal momentum distribution in \ref{p1x1plots}. There is excellent agreement between the hybrid simulation keeping up to mode 1 and the 3D simulation,  both quantitatively and qualitatively. 
   
\subsection{Beam Loading of a LWFA}
\label{beamloadingsection}
\begin{figure}[htbp] 
   \centering
   \includegraphics[width=0.5\textwidth]{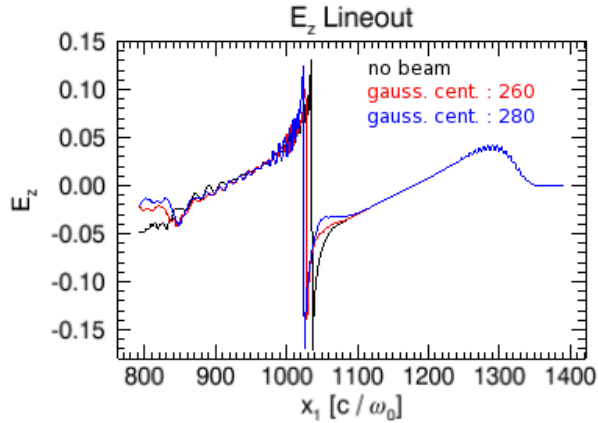} 
   \caption{Lineouts of $E_1$ along the laser for LWFA simulations including beam loading. The black is the wake without beam loading. The red and blue lines show the wake as it is loaded with a gaussian beam with its center at $z_0 = 260 c/\omega_0$, and $z_0=280 c/\omega_0$, respectively. The center of the laser is initially at $503 c/\omega_0$, and the laser has propagated $0.1mm$ into the plasma.}
   \label{beamloadingfig}
\end{figure}

Implementing the new algorithm into OSIRIS immediately provides the capability of modeling beam driven plasma based acceleration concepts as well as beam loading of laser produced wakes as well as modeling beam loading over pump depletion distances. We next show preliminary results where we beam load the wake in the LWFA simulations shown previously.   Recently, Tzoufras et al.\cite{beamloading} described how to analyze beam loading in nonlinear wakes, but there has been very little computational studies of beam loading in nonlinear wakes created by lasers due to the inability to routinely study this in three dimensions.  Here we show that the hybrid scheme could be a very useful tool for such studies. We loaded a gaussian beam with $k_p \sigma_z = 0.5$ and $k_p \sigma_r = 0.2$, and a peak density such that $(n_b / n_p) k_p^2 \sigma_r^2 \equiv \Lambda \approx 2$ into the wake.  The charge per unit length, $\Lambda$, is the critical normalized parameter which describes the degree of nonlinearity in the wakefield driven by the beam \cite{weilu, weilu2}. For $n_p = 1.5 \times 10^{18} cm ^{-3}$, this corresponds to a bunch with $\sigma_z = 2.2\mu m$, $\sigma_r = 0.87\mu m$, and $N \approx 1.9\times10^9 $ ($\approx300pC$). The spacing between the laser and the particle beam was varied. The trailing beam was initialized with an energy of 2 GeV, i.e., with a proper velocity of   $\gamma v_z = 40000.0 c$. The beam loading of the wake is presented in Figure~\ref{beamloadingfig}. This figure shows how the wake is loaded differently depending on the spacing between the laser and the trailing beam. In the future, the hybrid scheme will enable  routine studies of how the qualities of the trailing bunch and the overall efficiency depend on the location, shape, and current profile of the bunch. It will provide detailed parameter scans including the lowest order three dimensional effects and point towards parameters for full 3D simulations. 

\subsection{Hosing of Particle Beam Driver}
\label{hosingsection}
In this section we present an example of a particle beam driver, ie., PWFA. Axisymmetric $r$-$z$ simulations have been effectively utilized to study PWFA. However, such simulations cannot investigate asymmetric effects such as hosing and asymmetric spot-size effects.  Here we present a sample result for a PWFA simulation including $m \leq 2$ harmonics. The parameters are $n_b/n_p = 10.0$, with $k_p \sigma_z = 1$, and $k_p \sigma_r = 0.2$, $\Lambda = 0.4$. The simulation box size was $600c/\omega_p\times120c/\omega_p$ in the z and r directions, respectively. The initial beam proper velocity was $\gamma v_z = 40000.0 c$. The plasma was simulated with $16$ particles across $\phi$, while the beam was simulated with $32$ particles across $\phi$. There were $4$ particles per $r$-$z$ cell both species. The beam was initialized as an azimuthally symmetric beam.


\begin{figure}[htbp] 
   \centering
\begin{tabular}{c}
    \includegraphics[width=0.8\textwidth]{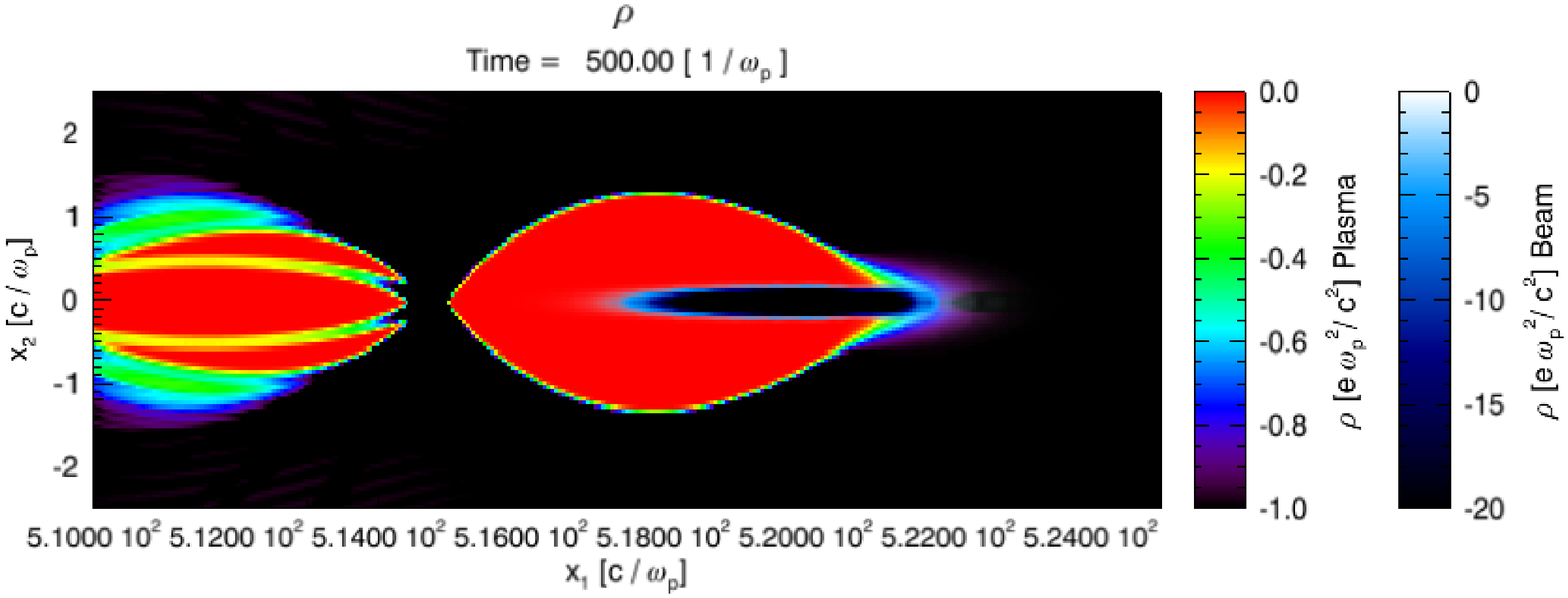}\\
   \includegraphics[width=0.8\textwidth]{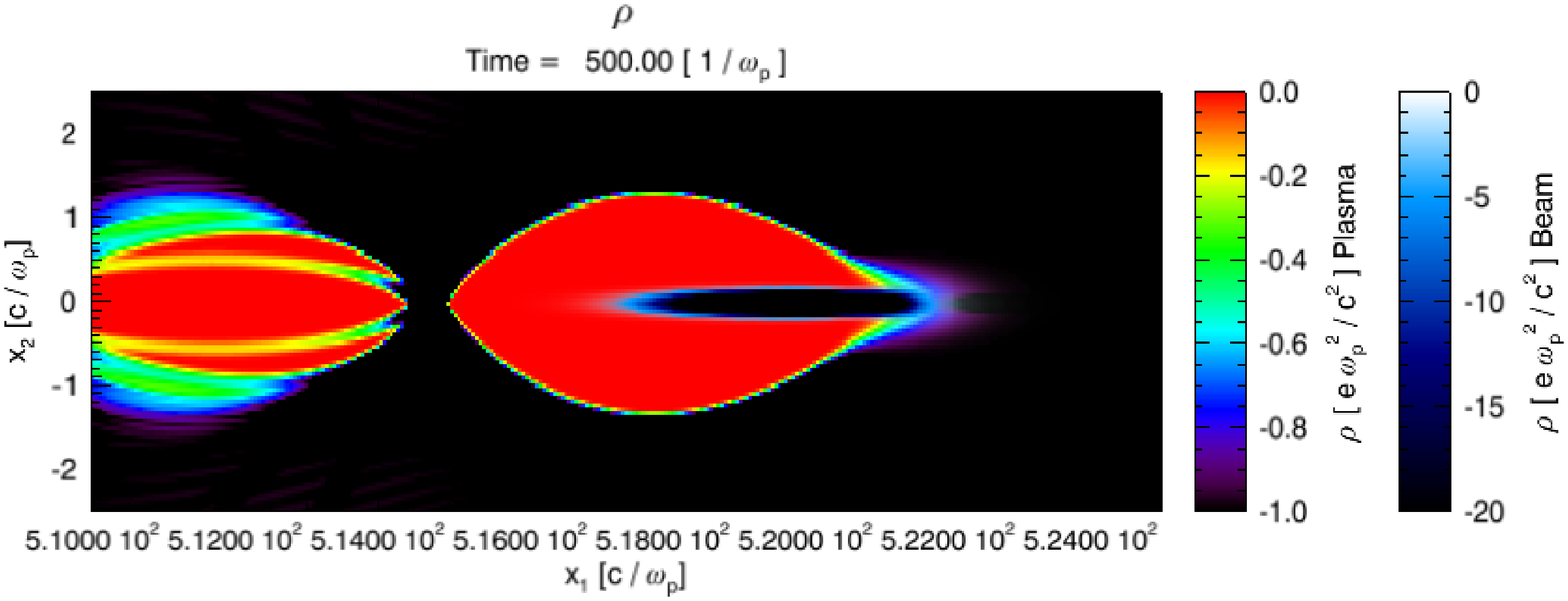} 
\end{tabular}
   \caption{The density from PWFA simulations with a 2D  cylindrical geometry simulation (top) and an $m \leq 2$ hybrid simulation (top). The beam has only moved $500 c / \omega_p$ into the plasma. The bottom plot was created by summing the modes at $\phi=0$ on the top half of the grid, and at $\phi = \pi$ on the bottom, which gives us the $y = 0$ cross-section of the three-dimensional beam. The 2D cylindrical simulation plot simply mirrors the bottom half from the top. For short distances the two simulations agree very well.}
   \label{beamcylcompare}
\end{figure}

\begin{figure}[htbp] 
   \centering
\begin{tabular}{c}
    \includegraphics[width=0.8\textwidth]{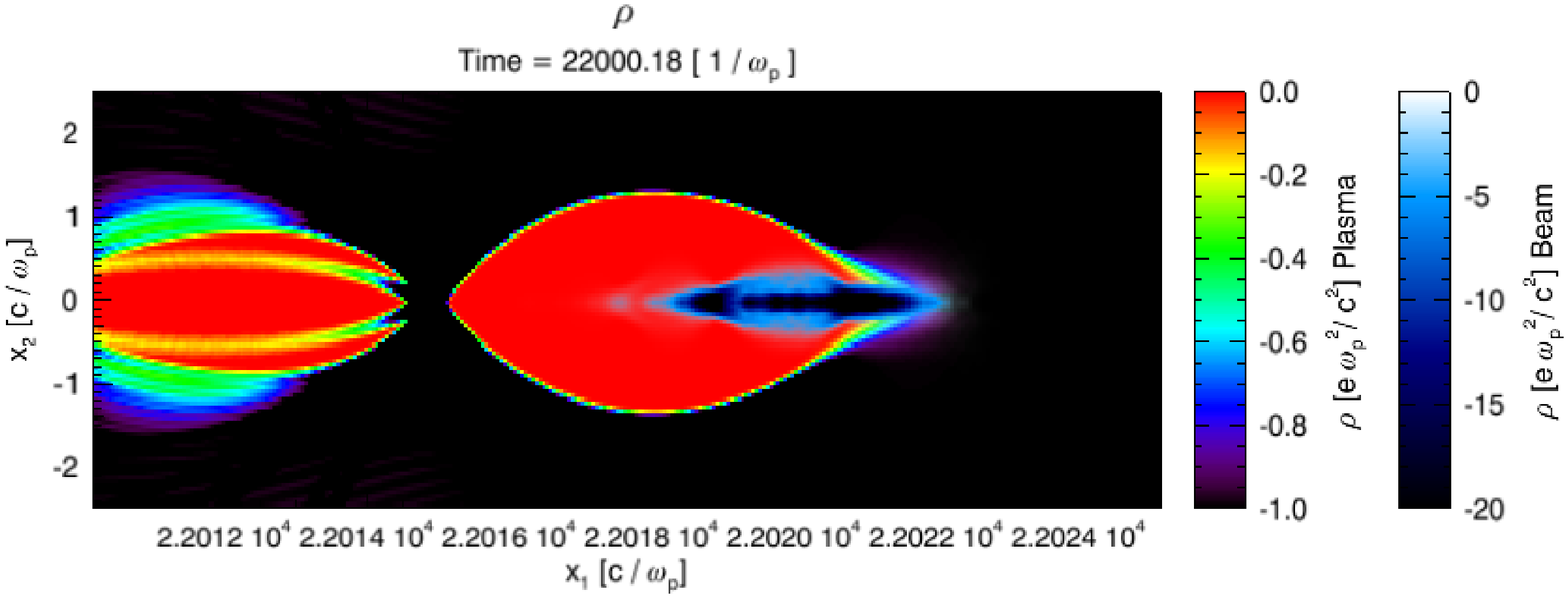}\\
   \includegraphics[width=0.8\textwidth]{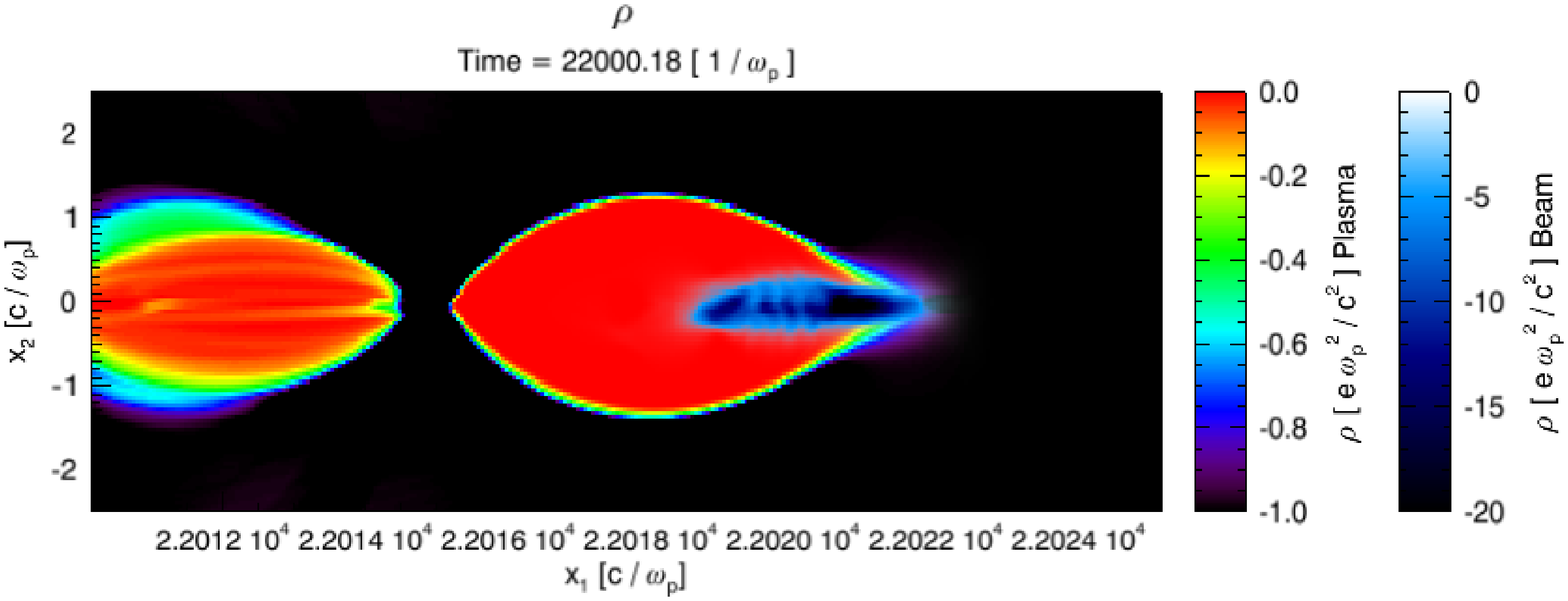} 
\end{tabular}
   \caption{The density plots after the beam has propagated $22000.18 c/\omega_p$ into the plasma. Hosing is observed $m \leq 2$ hybrid simulation (bottom). A result from an equivalent 2D cylindrical simulation is shown for comparison (top). The bottom plot was created by summing the modes at $\phi=0$ on the top half of the grid, and at $\phi = \pi$ on the bottom, which gives us the $y = 0$ cross-section of the three-dimensional beam. The 2D cylindrical simulation plot was generated by simply mirroring the bottom half from the top.}
   \label{beamhosing}
\end{figure}

\begin{figure}[htbp] 
   \centering
\begin{tabular}{c}
    \includegraphics[width=0.8\textwidth]{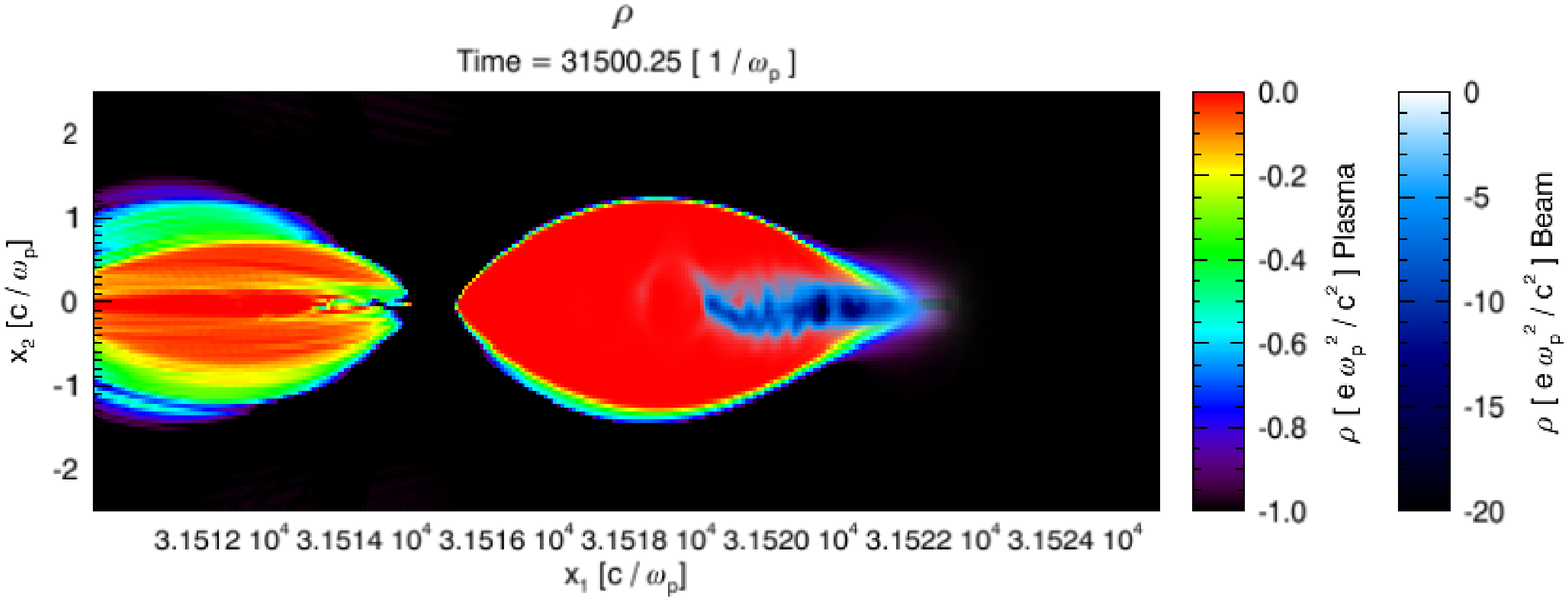}\\
   \includegraphics[width=0.8\textwidth]{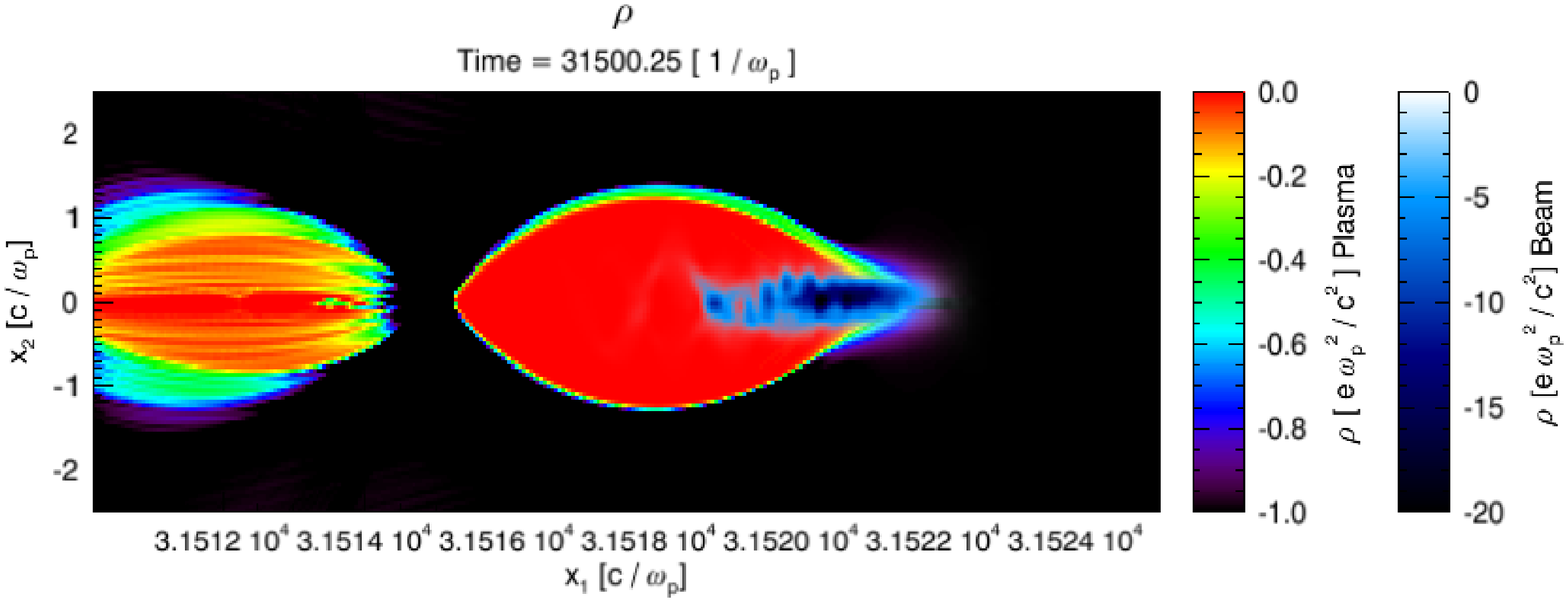} \\
\end{tabular}
\caption{Cross section of the beam and plasma density for the $x$-$z$ (top) and $y$-$z$ (bottom) planes. }
\label{asymspotsize}
\end{figure}


For this simulation we kept up to the $2^\text{nd}$ harmonic. Initially, as expected, the beam produces an azimuthally symmetric wake (and the beam remains symmetric). For short propagation distances the $m = 1$ and $m = 2$ modes are not important and the 2D r-z and hybrid 2D results look the same as seen in Figure~\ref{beamcylcompare}. The 2D r-z plot was generated by mirroring the result from positive $r$ to ``negative'' $r$. We also note that the 2D r-z code and full 3D results have been compared for round beams with no tilt and excellent agreement was found\cite{royphd}. For longer propagation distances hosing occurs. Hosing theory \cite{Huang07} is based on coupling the $m = 1$ modes for the centroid of the beam to that of the wake. In this simulation hosing grows from random noise in the beam density and it can clearly be seen in Figure~\ref{beamhosing}. The upper plot is from a 2D r-z simulation where hosing is precluded. The lower plot is from a 2D hybrid simulation. We emphasize that the result from a 2D hybrid and full 3D simulation will not quantitatively agree if physics which grows from a random noise source is important. However, if the dominant source for growth is the beam tilt agreement should exist and the growth rates should agree well. After a longer period of time the spot size begins to deviate due to a higher modal ($m = 2$) effect. This is shown in Figure~\ref{asymspotsize}.

\section{Conclusion}
\label{conclusion}

In this paper, we describe how we have implemented into OSIRIS the ability to expand the fields on an $r$-$z$ grid into an arbitrary number of Fourier harmonics in $\phi$. We used the fact that this is a hybrid PIC and gridless algorithm to develop a rigorous charge conserving current deposit for the hybrid algorithm. We showed that if the current amplitude for each harmonic in the $r$-$z$ plane is defined as $\mathbf{J}_{\perp,m} = \mathbf{J}_{\perp,0} e^{- i m \phi_p}$ where $\mathbf{J}_{\perp,0}$ is the current for the particle in the existing axisymmetric code and $\phi_p$ is the angle of the particle at the velocity time indicates, then an expression for $J_\phi$ on the grid can be derived that conserves charge for any particle order. We give examples that show the new scheme conserved charge to round off errors. We also present examples demonstrating the new algorithm's ability to efficiently study key physics in plasma based acceleration including LWFA, PWFA, and beam loading. 

The new algorithm reproduces qualitatively and quantitatively results from Lu et al. \cite{weilu} in the non-linear self-guided blowout regime for LWFA. The 3D Cartesian simulation requires $2 \times 300 = 600$ more cells ($2 \times N_z$ where $N_z$ is the number of cells in the transverse direction, and the factor of $2$ comes from only needing half the box in $r$) and $600 / 8 = 75$ more particles (the factor of $8$ comes from the number of particle empirically needed in $\phi$).

We also showed that keeping a few harmonics allows hosing of a particle beam to be studied and that both beam loading in laser driven wakes and hosing of the trailing beam can be studied by keeping only the $m = 0$ and $m = 1$ modes. The new code is currently $\approx 1/2$ the speed of the r-z code in a per particle basis when the $m = 1$ mode is included.  

Directions for future work include optimizing the algorithm to reduce the overhead of keeping $m$ copies of the mesh and interpolating the forces on the particles, including ionization, binary collision, and the PGC approximation, as well as additional field solvers with improved dispersion properties, and boundary conditions such as perfectly matched layers. We are interested in using the new hybrid code to study asymmetric spot size self-modulation and self focusing and other self-modulation processes \cite[and references therein]{mori97,duda, Sprangle94} for the laser and also how these couple to related instabilities for the trailing particles. We are also interested in using the new hybrid code to study laser solid interactions involved in fast ignition\cite{Tonge09,Fiuza11,May11} and proton acceleration\cite{Haberberger12}
%
, as well as stimulated Raman scattering \cite{Yin12,Winjum13} and the high frequency hybrid instability of a single speckle\cite{Afeyan95}. We will also pursue using this new scheme in a Lorentz boosted frame to obtain even more dramatic speed ups. 

This work was supported by the US Department of Energy contracts DE-SC0008491, DE-SC0008316, DE-NA0001833, DE-FC02-04ER54789, DE-FG02-92ER40727, US National Science Foundation under grants ACI 1339893,  the European Research Council (EU) through the Advanced Grant Accelerates (ERC-AdG2010 no 267841) and by EC FP7 through LaserLab-Europe/Laptech, and in China under the NSFC Grant 11175102, thousand young talents program.






\bibliographystyle{elsarticle-num}

\bibliography{davidson14}

\begin{thebibliography}{10}
\expandafter\ifx\csname url\endcsname\relax
  \def\url#1{\texttt{#1}}\fi
\expandafter\ifx\csname urlprefix\endcsname\relax\def\urlprefix{URL }\fi
\expandafter\ifx\csname href\endcsname\relax
  \def\href#1#2{#2} \def\path#1{#1}\fi

\bibitem{Lifschitz}
A.~Lifschitz, X.~Davone, E.~Lefebvre, J.~Faure, C.~Rechatin, V.~Malka,
  Particle-in-cell modelling of laser-plasma interaction using fourier
  decomposition, Journal of Computational Physics 228~(5) (2009) 1803--1814.
\newblock \href {http://dx.doi.org/http://dx.doi.org/10.1016/j.jcp.2008.11.017}
  {\path{doi:http://dx.doi.org/10.1016/j.jcp.2008.11.017}}.

\bibitem{Decker94}
C.~D. Decker, W.~B. Mori, T.~Katsouleas, {Particle-in-cell simulations of raman
  forward scattering from short-pulse high-intensity lasers}, {Physical Review
  E} {50}~({5}) ({1994}) {R3338--R3341}.

\bibitem{MoraAntonsen97}
P.~Mora, T.~Antonsen, {Kinetic modeling of intense, short laser pulses
  propagating in tenuous plasmas}, {Physics of Plasmas} {4}~({1}) ({1997})
  {217--229}.
\newblock \href {http://dx.doi.org/{10.1063/1.872134}}
  {\path{doi:{10.1063/1.872134}}}.

\bibitem{Huang06}
C.~Huang, V.~Decyk, M.~Zhou, W.~Lu, W.~Mori, Quicpic: a highly efficient fully
  parallelized pic code for plasma-based accelration, Journal of Physics:
  Conference Series 46 (2006) 190--199.
\newblock \href {http://dx.doi.org/10.1088/1742-6596/46/1/026}
  {\path{doi:10.1088/1742-6596/46/1/026}}.

\bibitem{an:13}
W.~An, V.~K. Decyk, W.~B. Mori, T.~M. Antonsen, Jr., {An improved iteration
  loop for the three dimensional quasi-static particle-in-cell algorithm:
  QuickPIC}, {Journal Of Computational Physics} {250} ({2013}) {165--177}.
\newblock \href {http://dx.doi.org/{10.1016/j.jcp.2013.05.020}}
  {\path{doi:{10.1016/j.jcp.2013.05.020}}}.

\bibitem{Gordon00}
D.~Gordon, W.~Mori, T.~Antonsen, {A ponderomotive guiding center
  particle-in-cell code for efficient modeling of laser-plasma interactions},
  {IEEE Transactions on Plasma Science} {28}~({4}) ({2000}) {1224--1232}.

\bibitem{Vay07}
J.~L. Vay, {Noninvariance of space- and time-scale ranges under a Lorentz
  transformation and the implications for the study of relativistic
  interactions}, {Physical Review Letters} {98}~({13}).
\newblock \href {http://dx.doi.org/{10.1103/PhysRevLett.98.130405}}
  {\path{doi:{10.1103/PhysRevLett.98.130405}}}.

\bibitem{Vay11}
J.~L. Vay, C.~G.~R. Geddes, E.~Cormier-Michel, D.~P. Grote, {Effects of
  hyperbolic rotation in Minkowski space on the modeling of plasma accelerators
  in a Lorentz boosted frame}, {Physics of Plasmas} {18}~({3}).
\newblock \href {http://dx.doi.org/{10.1063/1.3559483}}
  {\path{doi:{10.1063/1.3559483}}}.

\bibitem{Martins10}
S.~F. Martins, R.~A. Fonseca, W.~Lu, W.~B. Mori, L.~O. Silva, {Exploring
  laser-wakefield-accelerator regimes for near-term lasers using
  particle-in-cell simulation in Lorentz-boosted frames}, {Nature Physics}
  {6}~({4}) ({2010}) {311--316}.
\newblock \href {http://dx.doi.org/{10.1038/NPHYS1538}}
  {\path{doi:{10.1038/NPHYS1538}}}.

\bibitem{Martins09}
S.~F. Martins, R.~A. Fonseca, J.~Vieira, L.~O. Silva, W.~Lu, W.~B. Mori,
  {Modeling laser wakefield accelerator experiments with ultrafast
  particle-in-cell simulations in boosted frames}, {PHYSICS OF PLASMAS}
  {17}~({5}), {51st Annual Meeting of the Division-of-Plasma-Physics of the
  American-Physics-Society, Atlanta, GA, NOV 02-06, 2009}.
\newblock \href {http://dx.doi.org/{10.1063/1.3358139}}
  {\path{doi:{10.1063/1.3358139}}}.

\bibitem{xu13}
X.~Xu, P.~Yu, S.~F. Martins, F.~S. Tsung, V.~K. Decyk, J.~Vieira, R.~A.
  Fonseca, W.~Lu, L.~O. Silva, W.~B. Mori, {Numerical instability due to
  relativistic plasma drift in EM-PIC simulations}, {Computer Physics
  Communications} {184}~({11}) ({2013}) {2503--2514}.
\newblock \href {http://dx.doi.org/{10.1016/j.cpc.2013.07.003}}
  {\path{doi:{10.1016/j.cpc.2013.07.003}}}.

\bibitem{yu14}
P.~Yu, X.~Xu, V.~K. Decyk, W.~An, J.~Vieira, F.~S. Tsung, R.~A. Fonseca, W.~Lu,
  L.~O. Silva, Modeling of laser wakefield acceleration in lorentz boosted
  frame using em-pic code with spectral solver, Journal of Computational
  Physics 266~(1) (2014) 124--138.
\newblock \href {http://dx.doi.org/http://dx.doi.org/10.1016/j.jcp.2014.02.016}
  {\path{doi:http://dx.doi.org/10.1016/j.jcp.2014.02.016}}.

\bibitem{Godfrey13}
B.~B. Godfrey, J.-L. Vay, {Numerical stability of relativistic beam
  multidimensional PIC simulations employing the Esirkepov algorithm}, {Journal
  of Computational Physics} {248} ({2013}) {33--46}.
\newblock \href {http://dx.doi.org/{10.1016/j.jcp.2013.04.006}}
  {\path{doi:{10.1016/j.jcp.2013.04.006}}}.

\bibitem{Marder}
B.~Marder, {A Method For Incorporating Gauss Law Into Electromagnetic Pic
  Codes}, {Journal Of Computational Physics} {68}~({1}) ({1987}) {48--55}.
\newblock \href {http://dx.doi.org/{10.1016/0021-9991(87)90043-X}}
  {\path{doi:{10.1016/0021-9991(87)90043-X}}}.

\bibitem{Langdon92}
A.~B. Langdon, {On Enforcing Gauss Law In Electromagnetic Particle-In-Cell
  Codes}, {Computer Physics Communications} {70}~({3}) ({1992}) {447--450}.

\bibitem{birdsall1985plasma}
C.~Birdsall, A.~Langdon,
  \href{http://books.google.com/books?id=7TMbAQAAIAAJ}{Plasma physics via
  computer simulation}, The Adam Hilger series on plasma physics, McGraw-Hill,
  1985.
\newline\urlprefix\url{http://books.google.com/books?id=7TMbAQAAIAAJ}

\bibitem{ricardo}
R.~Fonseca, L.~O. Silva, F.~Tsung, V.~Decyk, W.~Lu, C.~Ren, W.~B. Mori,
  S.~Deng, S.~Lee, T.~Katsouleas, Osiris: A three-dimensional, fully
  relativistic particle in cell code for modeling plasma based accelerators,
  Lecture Notes in Computer Science 2331 (2002) 342--351.

\bibitem{dawson70}
J.~Dawson, {The electrostatic sheet model for a plasma and its modification to
  finite-size particles}, in: B.~Alder, S.~Fernbach, M.~Rotenberg (Eds.),
  {Methods in computational physics. IX. Plasma physics}, {Academic}, {London,
  UK}, {1970}, pp. {1--28}.

\bibitem{iprop}
B.~Godfrey, M.~R. C.~A. NM.,
  \href{http://books.google.com/books?id=hos\_OAAACAAJ}{The IPROP
  Three-Dimensional Beam Propagation Code}, Defense Technical Information
  Center, 1985.
\newline\urlprefix\url{http://books.google.com/books?id=hos\_OAAACAAJ}

\bibitem{ivory}
B.~Godfrey, M.~M. Campbell, M.~R. C.~A. NM., IVORY: A Three-Dimensional Beam
  Propagation Code, Defense Technical Information Center, 1982.

\bibitem{polaraxis}
G.~Constantinescu, S.~Lele, A highly accurate technique for the treatment of
  flow equations at the polar axis in cylindrical coordinates using series
  expansions, Computer Physics Communications 183~(1) (2002) 165--186.
\newblock \href {http://dx.doi.org/http://dx.doi.org/10.1006/jcph.2002.7187}
  {\path{doi:http://dx.doi.org/10.1006/jcph.2002.7187}}.

\bibitem{yee}
K.~Yee, Numerical solution of initial boundary value problems involving
  maxwell's equations in isotropic media, Antennas and Propagation, IEEE
  Transactions on 14~(3) (1966) 302--307.
\newblock \href {http://dx.doi.org/10.1109/TAP.1966.1138693}
  {\path{doi:10.1109/TAP.1966.1138693}}.

\bibitem{Esirkepov01}
T.~Esirkepov, {Exact charge conservation scheme for Particle-in-Cell simulation
  with an arbitrary form-factor}, {Computer Physics Communications} {135}~({2})
  ({2001}) {144--153}.
\newblock \href {http://dx.doi.org/{10.1016/S0010-4655(00)00228-9}}
  {\path{doi:{10.1016/S0010-4655(00)00228-9}}}.

\bibitem{chargecons}
J.~Villasenor, O.~Buneman, Rigorous charge conservation for local
  electromagnetic field solvers, Computer Physics Communications 69 (1992)
  306--316.
\newblock \href {http://dx.doi.org/10.1016/0010-4655(92)90169-Y}
  {\path{doi:10.1016/0010-4655(92)90169-Y}}.

\bibitem{weilu}
W.~Lu, M.~Tzoufras, C.~Joshi, F.~Tsung, W.~Mori, J.~Vieira, R.~Fonseca,
  L.~Silva, Generating multi-gev electron bunches using single stage laser
  wakefield acceleration in a 3d nonlinear regime, Physical Review Special
  Topics - Accelerators and Beams 10~(061301).
\newblock \href
  {http://dx.doi.org/http://link.aps.org/doi/10.1103/PhysRevSTAB.10.061301}
  {\path{doi:http://link.aps.org/doi/10.1103/PhysRevSTAB.10.061301}}.

\bibitem{beamloading}
M.~Tzoufras, W.~Lu, F.~Tsung, C.~Huang, W.~Mori, T.~Katsouleas, J.~Vieira,
  R.~Fonseca, L.~Silva, Beam loading by electrons in nonlinear plasma wakes,
  Physics of Plasmas 16~(056705).
\newblock \href {http://dx.doi.org/http://dx.doi.org/10.1063/1.3118628}
  {\path{doi:http://dx.doi.org/10.1063/1.3118628}}.

\bibitem{weilu2}
W.~Lu, C.~Huang, M.~Zhou, W.~B. Mori, T.~Katsouleas,
  \href{http://link.aps.org/doi/10.1103/PhysRevLett.96.165002}{Nonlinear theory
  for relativistic plasma wakefields in the blowout regime}, Phys. Rev. Lett.
  96 (2006) 165002.
\newblock \href {http://dx.doi.org/10.1103/PhysRevLett.96.165002}
  {\path{doi:10.1103/PhysRevLett.96.165002}}.
\newline\urlprefix\url{http://link.aps.org/doi/10.1103/PhysRevLett.96.165002}

\bibitem{royphd}
G.~Hemker, R, Particle-in-cell modeling of plasma-based accelerators in two and
  three dimensions, Ph.D. thesis, University of California, Los Angeles (2000).

\bibitem{Huang07}
C.~Huang, W.~Lu, M.~Zhou, C.~E. Clayton, C.~Joshi, W.~B. Mori, P.~Muggli,
  S.~Deng, E.~Oz, T.~Katsouleas, M.~J. Hogan, I.~Blumenfeld, F.~J. Decker,
  R.~Ischebeck, R.~H. Iverson, N.~A. Kirby, D.~Walz, {Hosing instability in the
  blow-out regime for plasma-wakefield acceleration}, {Physical Review Letters}
  {99}~({25}).
\newblock \href {http://dx.doi.org/{10.1103/PhysRevLett.99.255001}}
  {\path{doi:{10.1103/PhysRevLett.99.255001}}}.

\bibitem{mori97}
W.~Mori, {The physics of the nonlinear optics of plasmas at relativistic
  intensities for short-pulse lasers}, {IEEE Journal of Quantum Electronics}
  {33}~({11}) ({1997}) {1942--1953}.
\newblock \href {http://dx.doi.org/{10.1109/3.641309}}
  {\path{doi:{10.1109/3.641309}}}.

\bibitem{duda}
B.~Duda, W.~Mori, Variational principle approach to short-pulse laser-plasma
  interactions in three dimensions, Physical Review E 61~(2) (2000) 1925--1939.
\newblock \href {http://dx.doi.org/10.1103/PhysRevE.61.1925}
  {\path{doi:10.1103/PhysRevE.61.1925}}.

\bibitem{Sprangle94}
P.~Sprangle, J.~Krall, E.~Esarey, {Hose-Modulation Instability of Laser-Pulses
  In Plasmas}, {Physical Review Letters} {73}~({26}) ({1994}) {3544--3547}.
\newblock \href {http://dx.doi.org/{10.1103/PhysRevLett.73.3544}}
  {\path{doi:{10.1103/PhysRevLett.73.3544}}}.

\bibitem{Tonge09}
J.~Tonge, J.~May, W.~B. Mori, F.~Fiuza, S.~F. Martins, R.~A. Fonseca, L.~O.
  Silva, C.~Ren, {A simulation study of fast ignition with ultrahigh intensity
  lasers}, {Physics of Plasmas} {16}~({5}), {50th Annual Meeting of the
  Division of Plasma Physics of the American-Physical-Society, Dallas, TX, FEB
  01, 2008}.
\newblock \href {http://dx.doi.org/{10.1063/1.3124788}}
  {\path{doi:{10.1063/1.3124788}}}.

\bibitem{Fiuza11}
F.~Fiuza, M.~Marti, R.~A. Fonseca, L.~O. Silva, J.~Tonge, J.~May, W.~B. Mori,
  {Efficient modeling of laser-plasma interactions in high energy density
  scenarios}, {Plasma Physics and Controlled Fusion} {53}~({7}).
\newblock \href {http://dx.doi.org/{10.1088/0741-3335/53/7/074004}}
  {\path{doi:{10.1088/0741-3335/53/7/074004}}}.

\bibitem{May11}
J.~May, J.~Tonge, F.~Fiuza, R.~A. Fonseca, L.~O. Silva, C.~Ren, W.~B. Mori,
  {Mechanism of generating fast electrons by an intense laser at a steep
  overdense interface}, {Physical Review E} {84}~({2, 2}).
\newblock \href {http://dx.doi.org/{10.1103/PhysRevE.84.025401}}
  {\path{doi:{10.1103/PhysRevE.84.025401}}}.

\bibitem{Haberberger12}
D.~Haberberger, S.~Tochitsky, F.~Fiuza, C.~Gong, R.~A. Fonseca, L.~O. Silva,
  W.~B. Mori, C.~Joshi, {Collisionless shocks in laser-produced plasma generate
  monoenergetic high-energy proton beams}, {Nature Physics} {8}~({1}) ({2012})
  {95--99}.

\bibitem{Yin12}
L.~Yin, B.~J. Albright, H.~A. Rose, K.~J. Bowers, B.~Bergen, R.~K. Kirkwood,
  {Self-Organized Bursts of Coherent Stimulated Raman Scattering and Hot
  Electron Transport in Speckled Laser Plasma Media}, {Physical Review Letters}
  {108}~({24}).
\newblock \href {http://dx.doi.org/{10.1103/PhysRevLett.108.245004}}
  {\path{doi:{10.1103/PhysRevLett.108.245004}}}.

\bibitem{Winjum13}
B.~J. Winjum, J.~E. Fahlen, F.~S. Tsung, W.~B. Mori, {Anomalously Hot Electrons
  due to Rescatter of Stimulated Raman Scattering in the Kinetic Regime},
  {Physical Review Letters} {110}~({16}).
\newblock \href {http://dx.doi.org/{10.1103/PhysRevLett.110.165001}}
  {\path{doi:{10.1103/PhysRevLett.110.165001}}}.

\bibitem{Afeyan95}
B.~Afeyan, E.~Williams, {Unified Theory Of Stimulated Raman-Scattering And
  2-Plasmon Decay In Inhomogeneous Plasmas - High-Frequency Hybrid
  Instability}, {Physical Review Letters} {75}~({23}) ({1995}) {4218--4221}.
\newblock \href {http://dx.doi.org/{10.1103/PhysRevLett.75.4218}}
  {\path{doi:{10.1103/PhysRevLett.75.4218}}}.

\end{thebibliography}

\end{document}